\documentclass[conference, 10pt]{IEEEtran}
\IEEEoverridecommandlockouts

\usepackage{cite}
\usepackage{amsmath,amssymb,amsfonts}
\usepackage[ruled,linesnumbered]{algorithm2e}
\usepackage{graphicx}
\usepackage{textcomp}

\usepackage[normalem]{ulem}
\usepackage{xcolor}  %
\usepackage{enumitem}   %
\usepackage{lipsum}
\usepackage{booktabs}   %
\usepackage{pifont}     %
\usepackage{subfigure}  
\usepackage{xurl}  %
\usepackage{pifont}     %
\usepackage{circledsteps}   %
\usepackage{stfloats}   %
\usepackage{multirow} %
\usepackage{tabularx}
\usepackage{array} 

\usepackage{tikz}
\newcommand*\circled[1]{\tikz[baseline=(char.base)]{
  \node[shape=circle,draw,inner sep=1pt] (char) {\scriptsize #1};}}

\newcommand{\Section}{Section}  %

\newcommand{\revised}[1]{#1}
\newcommand{\revisednochange}[1]{#1}
\def\BibTeX{{\rm B\kern-.05em{\sc i\kern-.025em b}\kern-.08em
    T\kern-.1667em\lower.7ex\hbox{E}\kern-.125emX}}

\IEEEoverridecommandlockouts\IEEEpubid{\makebox[\columnwidth]{979-8-3315-0376-5/25/\$31.00 $\copyright$2025 IEEE \hfill}\hspace{\columnsep}\makebox[\columnwidth]{ }}

\begin{document}

\newcommand{\method}{INTA}

\title{\method{}: Intent-Based Translation for Network Configuration with LLM Agents}

\author{
	\IEEEauthorblockN{
		Yunze Wei\IEEEauthorrefmark{2}\thanks{This work was supported by NSFC Project under Grant 62132009 and Grant 62221003.}, 
		Xiaohui Xie\thanks{* Corresponding Authors: Xiaohui Xie and Yong Cui.}\IEEEauthorrefmark{2}\IEEEauthorrefmark{1}, 
		Tianshuo Hu\IEEEauthorrefmark{2}, 
		Yiwei Zuo\IEEEauthorrefmark{3}, 
		Xinyi Chen\IEEEauthorrefmark{2}, 
		Kaiwen Chi\IEEEauthorrefmark{2}, 
		Yong Cui\IEEEauthorrefmark{2}\IEEEauthorrefmark{1}} 
	\IEEEauthorblockA{\IEEEauthorrefmark{2}Department of Computer Science and Technology, Tsinghua University, Beijing, China}
	\IEEEauthorblockA{\IEEEauthorrefmark{3}Australian National University, Canberra, Australia}
}

\maketitle

\begin{abstract}

Translating configurations between different network devices is a common yet challenging task in modern network operations. 
This challenge arises in typical scenarios such as replacing obsolete hardware and adapting configurations to emerging paradigms like Software Defined Networking (SDN) and Network Function Virtualization (NFV).
\revisednochange{Engineers need to thoroughly understand both source and target configuration models, which requires considerable effort due to the complexity and evolving nature of these specifications.} %
To promote automation in network configuration translation, we propose \textbf{\method{}}, an intent-based translation framework that leverages Large Language Model (LLM) agents. 
The key idea of \method{} is to use \textit{configuration intent} as an intermediate representation for translation.
It first employs LLMs to decompose configuration files and extract fine-grained intents for each configuration fragment. 
These intents are then used to retrieve relevant manuals of the target device. %
Guided by a syntax checker, \method{} incrementally generates target configurations.
The translated configurations are further verified and refined for semantic consistency.
We implement \method{} and evaluate it on real-world configuration datasets from the industry.
Our approach outperforms state-of-the-art methods in translation accuracy and exhibits strong generalizability.
\method{} achieves an accuracy of 98.15\% in terms of both syntactic and view correctness, and a command recall rate of 84.72\% for the target configuration.
The semantic consistency report of the translated configuration further demonstrates its practical value in real-world network operations.

\end{abstract}


\section{Introduction}

Configuration translation has become an increasingly critical task in modern network operations and maintenance. 
As networks evolve, outdated or faulty devices are often replaced with more advanced and efficient alternatives~\cite{hartman2014equipment, zheng2024configtrans}, making configuration translation essential to ensure continuity and compatibility.
The adoption of Software Defined Networking (SDN) and Network Function Virtualization (NFV) further drives the need to integrate traditional network devices into SDN architectures or migrate their configurations to NFV environments~\cite{chen2022software, gsma2018migration}.
This often entails translating traditional command line interface (CLI) configurations into SDN controller-based or NFV-oriented representations.

Translating configurations across different network platforms is a complex and challenging task.
This paper takes cross-vendor CLI configuration translation as a representative example to illustrate the inherent difficulties of configuration translation.
Network engineers need to interpret the functionality and intent of complex source device configurations accurately~\cite{benson2009unraveling} and translate them into semantically equivalent configurations for target devices.
The vendor-specific nature of CLI syntax further complicates this process, requiring deep expertise in the configuration models of multiple vendor devices.
This requirement imposes substantial training costs on network engineers, as they must acquire in-depth knowledge of diverse vendor-specific configurations.
Such expertise is time-consuming to develop and difficult to maintain as network technologies and device models continue to evolve.

Both industry and academia are actively exploring automated methods for configuration translation.
NAssim~\cite{chen2022software} constructs device configuration models and uses NetBERT to recommend target commands. %
ConfigTrans~\cite{zheng2024configtrans} takes a step further by combining heuristic rules with Large Language Models (LLMs) to translate diverse types of commands. %
However, both approaches still require much manual work and struggle to generalize across diverse scenarios.
General programming language translation methods~\cite{roziere2020unsupervised, yang2024exploring} also fall short of meeting the requirements for this task due to the diversity of network device configuration syntax and the scarcity of configuration corpora.
We summarize the challenges of automated configuration translation into three core aspects:
(1) correctly understanding the logic and intent of the source configurations,
(2) accurately retrieving and interpreting the target device manuals, and
(3) generating configurations that are both syntactically correct and semantically consistent.

The rapid development of LLMs has brought novel opportunities for automated network configuration translation.
Recent successful applications of LLM-based multi-agent systems~\cite{hong2024metagpt, wu2024autogen} have demonstrated their potential for task understanding and solution generation. 
In this paper, we propose \method{}, an \underline{I}ntent-based framework for \underline{N}etwork configuration \underline{T}ranslation with LLM \underline{A}gents. 
Inspired by prior research on intent-driven approaches~\cite{jacobs2021hey,an2024nissist}, \method{} introduces intent as an intermediate representation that bridges the gap between cross-vendor device configurations.
\method{} comprises four key components: a \textit{parser} that parses the source configuration and extracts corresponding command manuals, an \textit{intent-based manual retriever} that analyzes configuration intents and retrieves relevant target-vendor device manuals through a two-stage retrieval and voting mechanism, a \textit{syntax-guided incremental translator} that incrementally generates the target configuration with syntactic guidance, and an \textit{LLM-based semantic verifier} that checks and refines the translation to ensure semantic consistency.
\method{} ultimately produces a target-device-specific configuration and a detailed report evaluating its syntactic correctness and semantic consistency, thereby assisting network engineers in verification and deployment.

We implement \method{} and evaluate it on a real-world configuration dataset collected from industry sources. 
In the router configuration translation scenario from Nokia to Huawei, \method{} achieved $98.15\%$ accuracy on view and syntactic correctness, and a command-level match rate of $84.72\%$ compared to reference translations.  
We also conduct ablation studies on multiple components of \method{} to validate their effectiveness.  
To provide deeper insight into \method{}'s workflow, we present a representative case study that demonstrates the end-to-end processing of a complete example.  
In addition, we validated \method{} in a switch scenario from Cisco to Huawei, achieving a syntactic correctness of $96.50\%$. 
These results demonstrate the effectiveness and generalizability of \method{} across different migration scenarios.
In addition, we manually verify the accuracy of the semantic consistency report, revealing \method{}'s practical value.

The main contributions of this paper are as follows:
\begin{enumerate}
\item We analyze the difficulties and challenges in cross-vendor network configuration translation, and propose \method{}, an intent-based configuration translation framework leveraging LLM agents to address these challenges.
\item We design an intent-based target device manual retrieval module to retrieve target device manuals accurately.
\item We develop a syntax-guided incremental translation module and a semantic refinement module to enhance the syntactic correctness and semantic consistency. %
\item We implement \method{} and evaluate it on real-world datasets across different network migration scenarios.
\end{enumerate}

\section{Background and Motivation} %

\subsection{Network Device Configuration}

Network device configuration is an essential part of network operation, covering the entire life cycle of network devices, including setup, maintenance, and troubleshooting.

There are several ways to configure network devices.
The most traditional and widely used method is the CLI, which requires administrators to write device configuration files and then manually input the configuration commands or automatically import these configuration files for deployment.
The NETCONF~\cite{rfc4741} protocol and the YANG~\cite{rfc6020, rfc6244} language are designed for advanced network device configuration and are widely used in SDN for data center or campus network operation.
Although new protocols provide more convenient possibilities for network configuration, the CLI is still indispensable in various scenarios, such as device initialization.
Given its complexity and widespread use, we focus on CLI configuration translation in this paper.

A CLI command on a network device typically comprises \textit{keywords}, \textit{parameters}, and the \textit{view} or operational context in which the command is executed.
A single CLI command line usually consists of \textit{keywords, parameters}, and many mandatory/optional items.
The following is an example of a command template for the Huawei NE40E Router~\cite{huawei:ip-address}.
\vspace{-1ex}
\begin{verbatim}
   ip address <ip-address> 
     { <mask> | <mask-length> } [ <sub> ]
\end{verbatim}
\vspace{-1ex}
In this template, \verb|ip address| is the keyword, \verb|<ip-address>| is the mandatory parameter,
\verb!{ <mask> | <mask-length> }! means that one of the parameters is required, and
\verb|[ <sub> ]| means that the parameter is optional.

A CLI often includes multiple \textit{views}, each containing a set of specific commands. For example, the command mentioned above may appear in the interface view, the Mtunnel view, and the ACL address pool view. 
Both the functionality and the parameters of the command can vary across different views.

\subsection{Configuration Translation}

\begin{figure}[t]
    \centering    
    \subfigure[One-to-many mapping.]{
    	\includegraphics[width=0.49\textwidth]{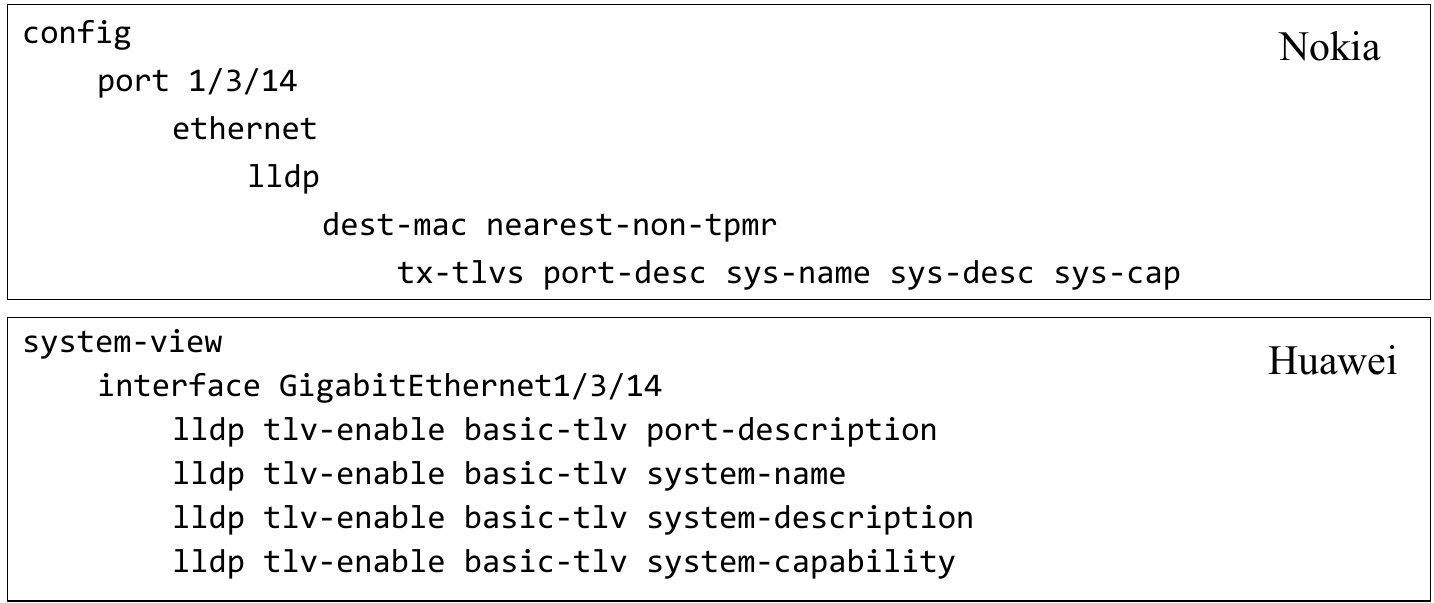}
    	\label{fig:1_to_N}
	}
    \subfigure[View depth differences.]{
        \includegraphics[width=0.49\textwidth]{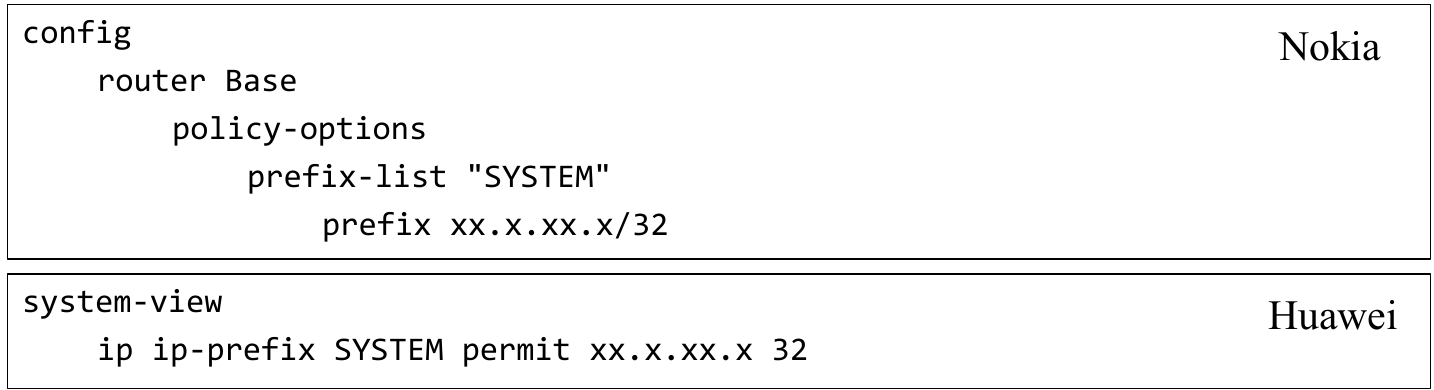}
        \label{fig:view_diff}
    }
    \subfigure[Device design logic differences.]{
        \includegraphics[width=0.49\textwidth]{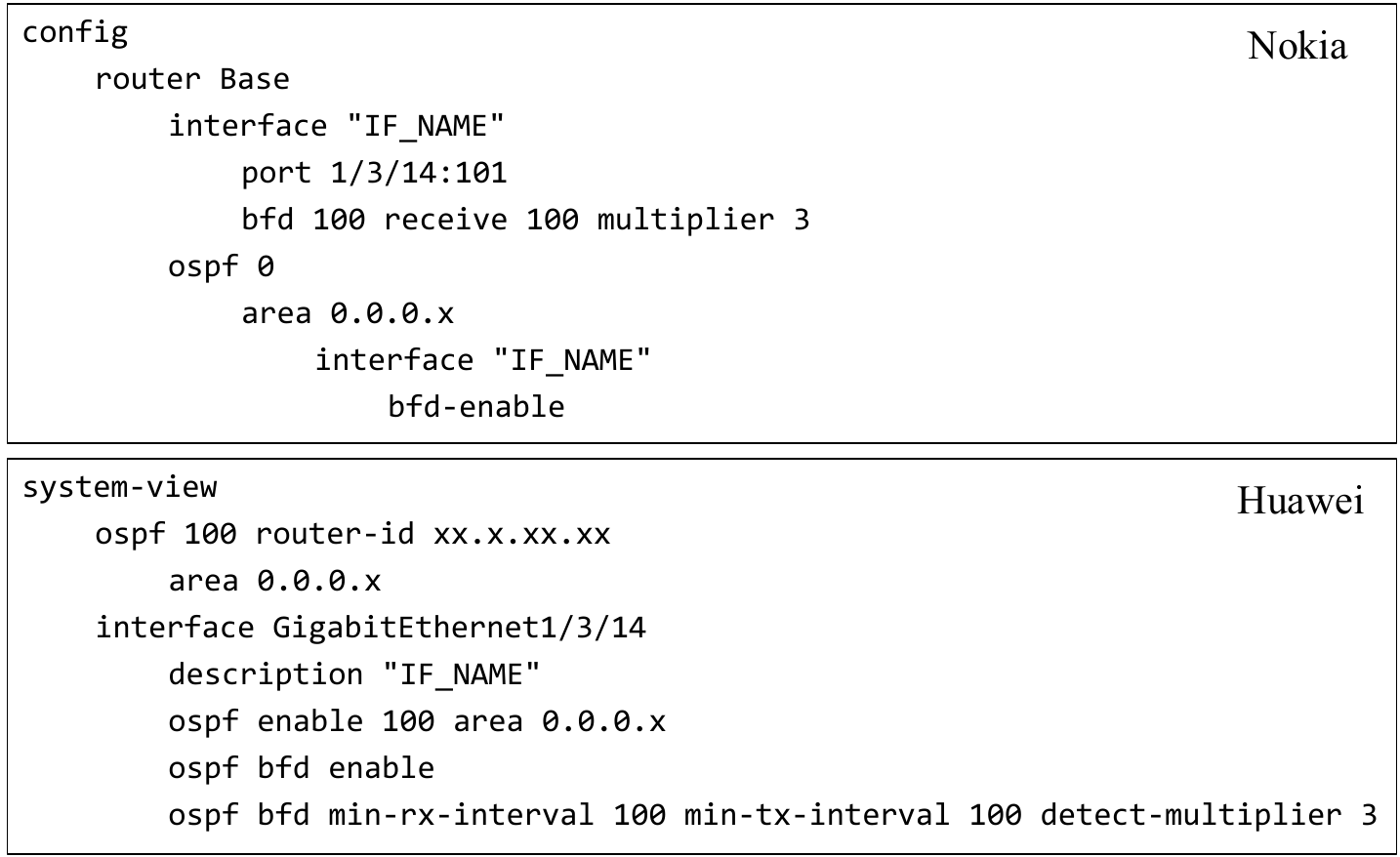}
        \label{fig:design_diff}
    }
    \caption{Design differences in different configuration models.}
    \vspace{-3ex}
    \label{fig:difficulties}
\end{figure}

\revisednochange{Network configuration translation is the process of converting a source device's configuration into that of a target device while preserving consistent behavior. 
This is often required when replacing devices from one vendor with another, typically driven by functionality upgrades, disaster recovery, policy changes, or cost considerations.}

The general process of configuration translation includes the following four steps:
(1) Understanding the intent and functionality of the source device configuration;
(2) Consulting device manuals to translate the source configuration into the corresponding target configuration;
(3) Analyzing the syntax and semantic correctness of the translated configuration; %
(4) Applying the translated configuration to the target device and monitoring the behavior of the target device.
This process is complex and demands considerable expert knowledge and experience.
Since the configuration models of different vendors vary significantly, translation requires experts who are familiar with both vendors' systems.  
We summarize the key differences in configuration models across vendors as follows.

\noindent \textbf{Design differences in configuration models.}
The difficulties of configuration translation stem from \textit{the significant design differences in different vendors' configuration models}, which can lead to substantial variations in configuration commands.
We summarize the differences in device configuration models into the following three aspects: one-to-many mappings, differences in view depth, and differences in design logic. 
We use the Huawei NE40E router and the Nokia 7750SR router as examples to illustrate these differences.

\textit{\circled{1} One-to-many mapping} is the simplest form of configuration differences. A typical case occurs when a single command in Nokia's device corresponds to multiple commands in Huawei's device, as shown in Fig.~\ref{fig:1_to_N}.

\textit{\circled{2} Differences in view depth} are the most apparent structural differences. As illustrated in Fig.~\ref{fig:view_diff}, configuring an IP prefix list in Huawei requires only a single command under the \texttt{system-view}, whereas in Nokia, the same task involves navigating through four levels of views.

\begin{figure}[t]
    \centering
    \setlength{\abovecaptionskip}{0ex}
    \includegraphics[width=0.47\textwidth]{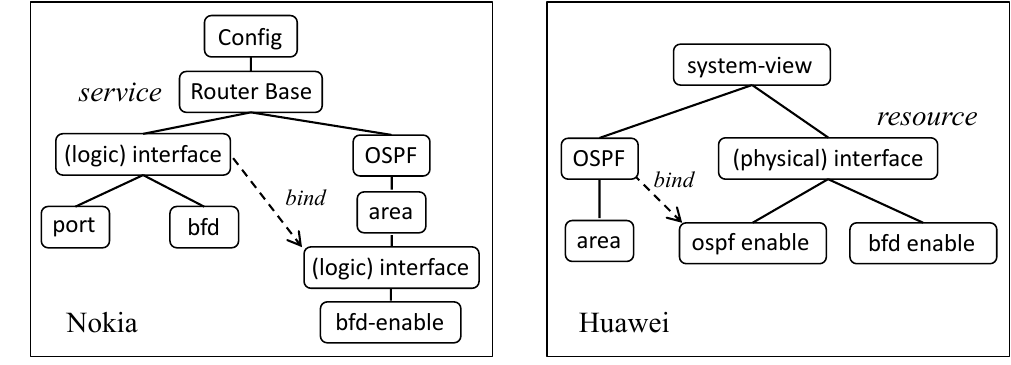}
    \caption{Configuration logic graph for Fig.~\ref{fig:design_diff}.}
    \label{fig:logic_graph}
    \vspace{-2ex}
\end{figure} 

\textit{\circled{3} Differences in design logic} are the fundamental reason why configuration translation is challenging. Vendors adopt distinct design principles for their configuration models, leading to substantial differences in configuration logic. For example, Fig.~\ref{fig:design_diff} demonstrates the configurations of OSPF and BFD protocols on Huawei and Nokia devices.
The configuration logic graphs are shown in Fig.~\ref{fig:logic_graph}.
Nokia adopts a \textit{service-centric}  approach.
It binds a pre-defined logical interface in the OSPF instance and enables the BFD protocol.
In contrast, Huawei adopts a \textit{resource-centric} approach.
It first creates an OSPF instance and then enables it in a physical interface, along with enabling the BFD protocol.
Accurately translating these configurations requires a deep understanding of the underlying design logic of each vendor's device model.

\begin{table}[t]
    \centering
    \caption{Comparison with existing methods.}%
    \resizebox{0.45\textwidth}{!}{%
    \begin{tabular}{lccc}
        \toprule
        \textbf{Method}                                  & \begin{tabular}[c]{@{}c@{}}\textbf{End-to-end}\\ \textbf{Translation}\end{tabular} & \begin{tabular}[c]{@{}c@{}}\textbf{Logic Difference}\\ \textbf{Understanding}\end{tabular} & \begin{tabular}[c]{@{}c@{}}\textbf{Migration}\\ \textbf{Overhead}\end{tabular} \\ 
        \midrule
        NAssim~\cite{chen2022software}          & \ding{55}   & \ding{55}   & Low \\
        ConfigTrans~\cite{zheng2024configtrans} & \ding{51}   & \ding{55}   & High \\
        \method{}                                    & \ding{51}   & \ding{51}   & Low \\
        \bottomrule
    \end{tabular}%
    }
    \label{tab:comparison} 
\end{table}
 
\noindent \revised{\textbf{Related work: mapping-based methods.}}
The most straightforward approach to configuration translation is directly mapping lines of configuration commands from one to another.
NAssim~\cite{chen2022software} constructs the device configuration model as a semantics-enhanced tree structure and uses NetBERT (fine-tuned SBERT~\cite{reimers-gurevych-2019-sentence}) to map the nodes of Vendor Device Model (VDM) and Unified Device Model (UDM) used in SDN.
\revised{However, NAssim is not an end-to-end solution. 
It still relies on human selection from the recommended commands.}
ConfigTrans~\cite{zheng2024configtrans} goes further based on NAssim's VDM models to achieve end-to-end cross-vendor translation.
It uses a heuristic method to translate commands with parameters, while commands without parameters are translated with the help of LLMs.
\revised{
However, ConfigTrans still relies on pre-defined rules and pre-built parameter correspondence tables, which require substantial manual work.}   %
Its heuristic algorithm also exhibits limited generalization across different configuration scenarios.

\revised{
In contrast to mapping-based methods, \method{} uses an intent-based method to bridge the gap between different configuration models.
As shown in Table~\ref{tab:comparison}, \method{} not only achieves end-to-end translation but also captures the underlying logical differences between configuration models, while maintaining low migration overhead across scenarios.}

\subsection{Challenges and Opportunities}\label{sec:challenges}

Based on an analysis of configuration translation methods in the industry, we identify three key challenges in the translation process and explore how LLM-based approaches offer promising opportunities to address them.

\noindent \textbf{C1: Interpreting the logic and intent of source configurations.}
A fundamental challenge in configuration translation is accurately interpreting the logic and intent behind the source configuration commands. While functional descriptions of individual commands can typically be found in command manuals, synthesizing these descriptions into a coherent, high-level intent representation remains difficult.

\noindent \textbf{C2: Retrieving relevant target device manuals.}
Generating an equivalent configuration requires identifying which commands on the target device can fulfill the intent, as well as determining their exact syntax (e.g., keywords, parameters, hierarchical structure). 
This necessitates retrieving relevant information from the target device's manuals, which is challenging due to the large volume of the manual corpus and the often vague or abstract nature of the configuration intent.

\noindent \textbf{C3: Generating syntactically valid and semantically consistent configurations.} 
The translated configuration must strictly adhere to the target device's complex syntax rules while preserving the original semantics of the source configuration. Achieving both syntactic correctness and semantic equivalence is non-trivial and requires carefully designed mechanisms.

\noindent \textbf{Opportunities of LLM-based approaches.} 
The significant advances in LLMs' understanding and reasoning abilities~\cite{huang2023towards} open new possibilities for configuration translation. Specifically, our proposed method, \method{}, uses LLMs to analyze configuration intent in place of human engineers (C1), assist in the retrieval of relevant manual content (C2), and perform incremental translation under the guidance of syntax checkers, followed by semantic verification and refinement (C3).

\section{Design of \method{}} %
\begin{figure*}[!htbp]
    \centering
    \includegraphics[width=0.99\textwidth]{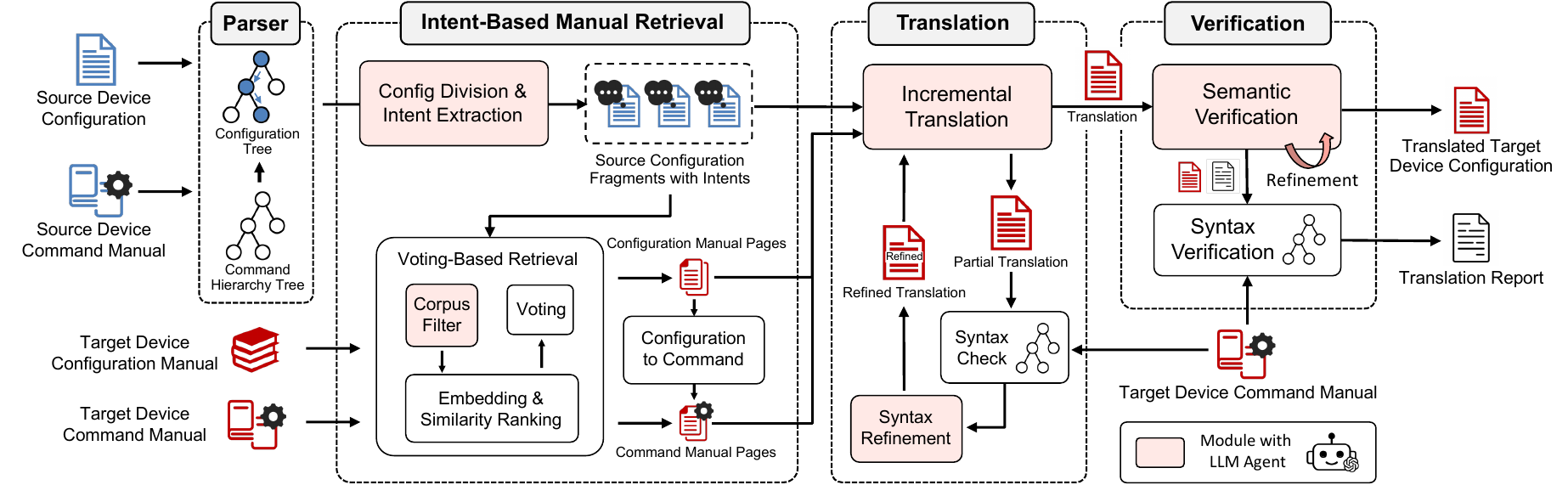}
    \caption{System workflow of \method{}.}
    \vspace{-1ex}
    \label{fig:system_overview}
\end{figure*}
\subsection{System Overview}

The workflow of \method{} is shown in Fig. \ref{fig:system_overview}, which consists of four main components: the configuration \textit{parser}, 
the intent-based \textit{retrieval} module for target device manuals,
the configuration \textit{translation} module, and the \textit{verification} module.

\noindent \textbf{Workflow.}
The source device configuration to be translated is first parsed by the configuration \textit{parser}.
The main component of the parser is a command hierarchy tree constructed from the command manuals of the source device and VDM, similar to NAssim~\cite{chen2022software}.
It parses the syntax of each line of configuration and maps the line on the command hierarchy tree, obtaining each command's view structure and corresponding manuals.
The details of the configuration parser are shown in Appendix~\ref{appdx:parser-details}.
The parsed configuration commands (together with the corresponding command manual pages) then enter the \textit{intent-based retrieval} module, which first divides the configuration into fragments and analyzes the intents of each fragment.
Then the configuration fragments with intents are used to retrieve the target device manuals, as detailed in \Section{}~\ref{sec:IR}.
The \textit{translation} module uses the retrieved configuration and the target device manual pages to translate the source configuration fragments.
This is an incremental process where each fragment is translated based on the preceding translation.
We also perform syntax checks and refinements to enhance this process.
Finally, the \textit{verification} module makes semantic verification and refinement, eliminating redundant information and supplementing missing semantic details.
The final output consists of the translated configuration and an accompanying translation report documenting syntax correctness and semantic consistency.
\revised{The overall workflow employs multiple LLM agents to address simpler, well-defined subtasks, leading to improved performance over  end-to-end approaches.}  %

\noindent \textbf{Two types of device manuals.} 
We use two types of manuals.
\textit{Configuration manuals} describe the procedures to implement specific functions, indicating which commands to use.
\textit{Command manuals} provide comprehensive syntax definitions and functional descriptions for all the commands, which also serve as the basis for constructing command hierarchy trees.
Examples of both manuals are shown in Appendix~\ref{sec:manual_examples}.

\subsection{Intent-Based Target Manuals Retrieval}\label{sec:IR}

\revisednochange{The retrieval of target device configuration and command manuals is essential for high-quality translation, as general LLMs lack domain-specific knowledge and thus require external manual injection.
However, accurately retrieving these manuals is a challenging task.}
Inspired by existing works~\cite{jacobs2021hey, an2024nissist}, we leverage configuration intent to bridge the significant gap between configuration and manuals.
\revised{In this process, intent serves as a vendor-neutral abstraction layer for manual retrieval and configuration translation, which is better suited for LLMs to understand and process.
To the best of our knowledge, we are the first to use ``intent'' as an intermediate representation for configuration translation. 
} %
We first use LLM to split the source configuration into fragments based on functionality and extract the intent of each fragment. %
Then we retrieve the corresponding manual pages of the target device based on the intent. %
We further use the retrieved configuration manuals to enhance the retrieval of command manuals. %

\subsubsection{Configuration Intent Extraction}\label{sec:intent_extraction}

We employ LLM to split the source configuration into fragments,
and extract the intent of each fragment.
\revised{The fragments are divided by LLMs based on functionality units inferred from the semantics of the source configuration and its corresponding command manual pages.}
To produce stylistically consistent and structured intent descriptions that support reliable manual retrieval, we adopt the In-Context Learning (ICL) method~\cite{dong2022survey, bertsch2024context}, using templates and examples to guide the LLM's intent extraction process.
We also ask the LLM to extract intents at different levels: a \textit{general} description of the entire configuration fragment and \textit{detailed} descriptions of each sub-module in the fragment.
This helps to improve the recall rate of the subsequent manual retrieval step.
\revised{The prompt template includes output requirements, few-shot examples, configuration to divide, corresponding source device manuals, etc., as shown in Fig.~\ref{fig:IRAG-ICL} (Appendix~\ref{sec:prompts}).}

\subsubsection{Voting-Based Target Manual Retrieval}\label{sec:target_manual_retrieval}

\begin{figure}
    \centering
    \includegraphics[width=0.43\textwidth]{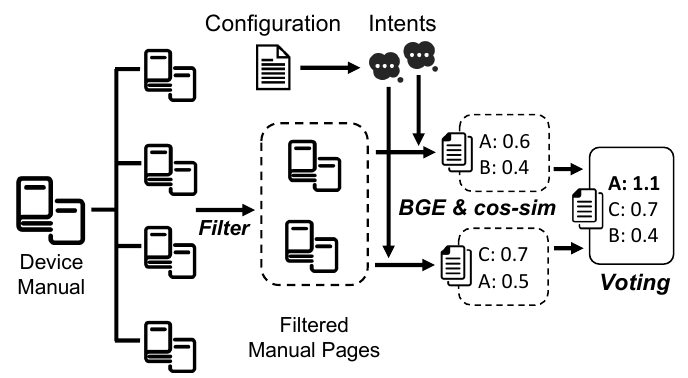}
    \caption{Detailed workflow of voting-based retrieval.}
    \vspace{-1.5ex}
    \label{fig:irag-manual}
\end{figure}

The detailed workflow of the proposed retrieval pipeline is shown in Fig.~\ref{fig:irag-manual}, which consists of three main steps: manual corpus filtering, manual context embedding, and voting mechanism.

\noindent \textbf{Manual corpus filtering.}
\revisednochange{The configuration and command manuals of network devices are massive. For instance, the Huawei NE40E router has a configuration manual of $\sim$8000 pages~\cite{huawei:config-manual} and a command manual of $\sim$14000 pages~\cite{huawei:cmd-manual}, making accurate retrieval particularly challenging. To address this, an initial filtering stage is essential to narrow the corpus, improving retrieval accuracy and efficiency.}
\revisednochange{Considering the distinct characteristics of configuration and command manuals, we adopt two separate filtering strategies.}

\textit{\circled{1} Filter for configuration manual corpus.}
The configuration manual is organized as a hierarchical directory tree, with each directory level clearly described using natural language.
Inspired by the way humans navigate and comprehend manuals, we propose to employ LLMs to interpret the directory and perform effective manual filtering.
We provide the LLM with the source configuration fragment, its corresponding manual content, and the directory structure of the target device's configuration manual, allowing LLM to select the most relevant directory.
However, high-level directory entries are often too coarse-grained (e.g., \verb|IP Service| and \verb|IP Routing|), making it difficult for the LLM to determine the appropriate directory accurately.
Therefore, we concatenate the first and second-level directories together, allowing LLM to obtain information from both levels simultaneously, enabling it to make more accurate selections. 
\revised{
The prompt template for LLM manual filtering is shown in Fig.~\ref{fig:IRAG-LLM-Filter} (Appendix~\ref{sec:prompts}), which includes the task and guidelines, source configuration with manuals, target configuration manual list, etc.}

\textit{\circled{2} Filter for command manual corpus.}
\revised{We use BM25~\cite{robertson2009probabilistic}, a classic probabilistic ranking function that scores documents based on term frequency and document length normalization, to perform initial filtering on the command manual.}
Experiments show that using BM25 and LLM-filter on command manuals yields similar performance (Section~\ref{sec:ra-module}), but BM25 is more cost-efficient and faster.

\noindent \textbf{Manual context embedding.}
\revisednochange{Manual pages typically include brief descriptions that outline their functionality.
We aim to match the extracted intents with these descriptions to identify the corresponding manual pages.}
We use the BGE model~\cite{bge-m3} based on BERT to encode intent sentences and manual contexts.
To disambiguate commands that may appear in multiple views, we augment each context with its manual file path, which provides view-specific information and helps filter out irrelevant views.
To sum up, the manual context we use includes: the title of a manual page, the page description, and the manual file path.
We also include CLI commands in the context for retrieving command manual pages.
\revisednochange{Finally, we compute the cosine similarity between the intent description and each manual context embedding.
The top-$k$ manuals with the highest scores are selected as retrieval results for the corresponding intent.}
 
\noindent \textbf{Voting mechanism.}
The voting mechanism is a crucial strategy for enhancing the recall rate of relevant target device manuals.
After the preceding processing steps, we obtain configuration fragments, their corresponding intent descriptions (both at the fragment level and for each individual command), and a top-k list of manuals retrieved for each intent.
To further improve the relevance of the retrieval results, we adopt a voting-based integration approach.
Specifically, for each intent description, we aggregate the retrieved manual lists using a weighted voting scheme, where each manual's similarity score serves as its voting weight.
The final score of each manual is calculated as the cumulative sum of its scores across all individual retrieved manual lists.
The manuals are then re-ranked based on their aggregated scores, and the resulting list serves as the knowledge base for the subsequent configuration translation stage.
This method can effectively improve the manual recall rate (\Section{}~\ref{sec:ra-module}).

\begin{algorithm}[t]
    \caption{Configuration-to-Command Retrieval}
    \label{algo:c2c-retrieval}
    \KwData{Retrieved configuration manual pages with scores $M = \{m_i \mapsto s_i\}$}
    \KwResult{Aggregated command manual pages with scores $M' = \{m'_j \mapsto s'_j\}$}
    
    \BlankLine
    $M' \leftarrow \{\}$\;
    
    \ForEach{$m \mapsto s \in M$}{
        $C \leftarrow \text{extract\_configuration\_commands}(m)$ %

        \ForEach{$c \in C$}{
            $m' \leftarrow \text{get\_command\_manual}(c)$
            
            $M'[m'] \leftarrow M'[m'] + s$ \tcp*{Default to 0 if $m' \notin M'$}
        }
    }
    \Return{$M'$}
\end{algorithm}

\subsubsection{Configuration-to-Command Manual Retrieval}\label{sec:c2c_manual_retrieval}

To further enhance the retrieval performance of command manuals, we adopt a cross-retrieval approach.
\revised{Specifically, we leverage command references in the previously retrieved configuration manual pages to supplement and refine the command manual retrieval.}
We use Algorithm~\ref{algo:c2c-retrieval} to map configuration manual pages to command manual pages.
The input to the algorithm is a dictionary mapping each configuration manual page $m$ to its retrieval score $s$. 
The output is a set of command manuals retrieved based on these configuration manuals.
The algorithm iterates over each configuration manual page $m \in M$, and uses the rule-based \verb|extract_configuration_commands| method to extract the configuration commands referenced by $m$ (line 3).
For each extracted command $c \in C$, it then applies the \verb|get_command_manual| method, which leverages an automatically constructed mapping from the command manual, to retrieve the corresponding command manual page $m'$ (line 5).
The scores of all retrieved command manual pages are aggregated and stored in $M'$ (line 6).
Finally, the command manual pages obtained through this process are merged with the previously retrieved list. Their scores are then aggregated and re-ranked to yield the final set of command manual pages.

\subsection{Syntax Checker Guided Incremental Translation}\label{sec:translation}

We incrementally translate configuration fragments from the source device into the target device's configuration based on the previously retrieved target device manuals, with the translation process guided by the command hierarchy tree.

\subsubsection{Incremental Translation}

Incremental translation refers to the sequential translation process of source configuration fragment by fragment.
We construct a comprehensive prompt by combining source device configuration commands, corresponding manuals, and previously retrieved target device configuration and command manuals.
Then we employ LLMs' understanding, analysis, and generation capabilities for translation.
\revisednochange{Since LLMs have context length limits~\cite{hsieh2024ruler} and excessively long contexts may degrade output quality~\cite{li2024long},}
we opted for a fragment-by-fragment rather than full-text translation approach.
While it is possible to translate each fragment independently, we observe that configuration fragments often exhibit strong interdependencies in practice.
\revised{Therefore, we adopt an incremental translation approach that leverages the translations of preceding fragments as context, allowing forward dependencies to be preserved across fragments.}
\revised{The prompt template includes manual command conventions, source commands with manuals, retrieved target manuals, etc., as shown in Fig.~\ref{fig:IRAG-Translation} (Appendix~\ref{sec:prompts}).}

\subsubsection{Syntax Checker Guidance}

This enables real-time checking of both syntax and view structure, as well as providing refinement hints for the incremental translation results.
Due to the hallucination phenomenon inherent in LLMs~\cite{huang2025survey}, the translated configuration commands may not conform to the configuration views or syntactic requirements of the target device.
Therefore, we build the command syntax and hierarchy checker using the target device's command manual and the VDM~\cite{chen2022software}, similar to the configuration parser module (Appendix~\ref{appdx:parser-details}).
We conduct two rounds of checks on the translated configuration: a view consistency check and a syntactic validity check.
In the first round, the translated configuration is validated against the command hierarchy tree, which reflects both the correct view and syntax constraints. Unmatched entries from this round may result from either view inconsistencies or syntax errors.
To distinguish the cause, a second round of matching is performed against the complete set of configuration commands, ignoring view constraints and focusing solely on syntax validity. Entries that remain unmatched are identified as \textit{syntax errors}, while those that match the full command set but not the hierarchy tree are labeled as \textit{view errors}.
We annotate the two types of errors and guide the LLM to correct the translated configuration from the previous iteration through multi-round dialogue, where the translation history is recorded during the process.
If these corrections lead to improved syntactic and view accuracy, the updated commands are adopted.
\revised{The prompt template for multi-round syntax correction is shown in Fig.~\ref{fig:ver-syntax} (Appendix~\ref{sec:prompts}).}

\subsection{Verification}\label{sec:verification}
\revisednochange{In the final stage of configuration translation, we perform semantic consistency verification and refinement on the translated configuration.
Furthermore, syntactic verification of the final translated configuration is also included.
The final translation output is accompanied by a verification report on semantic consistency and syntactic correctness, which facilitates better understanding and utilization by network operators.}

\subsubsection{Semantic Verification and Refinement}

\begin{algorithm}[t]
    \caption{Semantic Verification and Refinement}
    \label{algo:semantic}
    \KwData{Source configuration $s$, Translated configuration $t$, Source device manuals $M_s$, Target device manuals $M_t$}
    \KwResult{Refined configuration $t'$, Semantic verification report $r_1$}
    
    \BlankLine
    $r_0 \leftarrow \text{LLM\_semantic\_analysis}(s, t)$ %

    $t' \leftarrow t$\;
    
    \ForEach{$(s_i, t_i, \text{is\_consistent}, cmt) \in r_0$}{
        \If{$\text{is\_consistent} = \text{False}$}{
            $m_s \leftarrow \text{retrieve\_relevant\_manuals}(s_i, M_s)$\;
            $m_t \leftarrow \text{retrieve\_relevant\_manuals}(t_i, M_t)$\;
            $t'' \leftarrow \text{LLM\_semantic\_refinement}(s, t, s_i, t_i, m_s, m_t, cmt)$ %

            \If{$\text{syntax\_errors}(t'') \le \text{syntax\_errors}(t')$}{
                $t' \leftarrow t''$
            }
        }
    }
    
    $r_1 \leftarrow \text{LLM\_semantic\_analysis}(s, t')$ %
    
    \Return{$t', r_1$}
\end{algorithm}

The purpose of semantic verification is to analyze whether the translated configuration is semantically equivalent to the original source configuration. 
We observe in practice that LLMs excel at analyzing cross-vendor configuration differences, so we leverage them as semantic verifiers and refiners.
The process of semantic verification and refinement is illustrated in Algorithm~\ref{algo:semantic}.
The algorithm begins with an LLM-based analysis of semantic consistency between the translated and source configurations (line 1).
The output of this step is an initial semantic consistency report $r_0$, composed of multiple report units. Each unit contains a translated configuration fragment $t_i$, its corresponding source fragment $s_i$, a consistency flag $is\_consistent$, and an explanatory comment $cmt$ generated by the LLM.
We then iterate through all units in the report. For each unit marked as inconsistent, we perform a targeted refinement of the translation.
This refinement is guided by the corresponding pages from the source and target device manuals, the source and translated configuration fragments, and the LLM-generated comment (line 7).
After all inconsistent fragments have been refined, we conduct a second round of semantic analysis using the LLM to generate the final semantic consistency report $r_1$ (line 13).
The algorithm ultimately returns the semantically refined translation $t'$ along with its corresponding consistency report.
\revised{The prompt templates for semantic verification and refinement are shown in Fig.~\ref{fig:ver-semantic-prompt-template} and Fig.~\ref{fig:ver-semantic-refinement}, respectively, within Appendix~\ref{sec:prompts}.}

\subsubsection{Syntax Verification}
After semantic refinement, the translated configuration undergoes a final round of syntax verification.
\revisednochange{This process employs the same syntax checker introduced in Section~\ref{sec:translation}, where the command hierarchy tree is constructed based on the target device's command manual and its corresponding VDM.}
The translated configuration is then validated through line-by-line matching against this hierarchy tree.
\revisednochange{Notably, syntax verification checks both view hierarchy correctness and command syntax validity.}
The output of syntax verification includes the mapping between each configuration line and its corresponding command template.
Commands that fail to match are labeled as \verb|Mismatch|.

\subsubsection{Translation Report}

The translation report provides a comprehensive evaluation of translation output, focusing on syntactic correctness and semantic consistency.
Its primary goal is to assist human experts in efficiently leveraging the translation results.
While \method{} achieves high translation accuracy, it cannot fully ensure compliance with device-specific syntax or guarantee semantic equivalence.
The report mitigates these limitations by explicitly highlighting syntactic errors and potential semantic divergences, thereby reducing the verification burden for network engineers.
Since the syntax correctness report is generated by a deterministic configuration syntax checker, its findings are theoretically sound.
Similarly, the semantic consistency report identifies deviations between the source and translated configurations.
\revisednochange{While LLM-generated annotations occasionally contain errors,}
experimental results (\Section{}~\ref{sec:report}) demonstrate that the report remains reliable and practically useful.

\section{Evaluation and analysis} %

We implement \method{} in Python with $\sim$3500 lines of code (LoC).  %
For the parser, we build the configuration tree based on the open-source configuration models from NAssim~\cite{chen2022software} and implement vendor-specific parsers (Appendix~\ref{appdx:parser-details}).
In the intent-based manual retrieval module, we use BGE-M3~\cite{bge-m3} as the sentence embedding model.
The embedding process is accelerated by an NVIDIA GeForce RTX 3090 or A100 GPU.
\revised{
We use OpenAI's SDK to call several LLMs, including GPT-4o~\cite{openai:gpt-4o}, Qwen-Max~\cite{alibaba:qwen-max}, and DeepSeek-V3~\cite{liu2024deepseek} (671B).}\footnote{\revised{Specific model versions: gpt-4o-2024-08-06, qwen-max-2025-01-25,  deepseek-v3-0324.}}
We use a Huawei NE40E router in the GNS3 emulator~\cite{gns3} and a physical Huawei CE6881 switch for translation validation.
Our system has relatively low migration overhead.
Supporting a new vendor only requires scraping its manuals and modifying $\sim$100 LoC in the vendor-specific parser, without fine-tuning. 
The command syntax parser is reusable across vendors.

\subsection{Experimental Setup}

\noindent \textbf{Manuals and dataset.}
We use Nokia 7750 SR~\cite{nokia:7750SR} and Huawei NE40E~\cite{huawei:ne40e} routers as our source and target devices for our main experiments.
The command manuals and VDM (hierarchy of command manuals) come from the open-source dataset from NAssim~\cite{chen2022software}.
We scrape the configuration manual of NE40E from the Huawei website~\cite{huawei:config-manual}. %
To support full-coverage configuration translation, we retrieve content from the complete set of manuals, rather than a small subset as in ConfigTrans~\cite{zheng2024configtrans}.
This enables broader applicability but also poses a much greater challenge for accurate manual retrieval.
Our dataset includes $1063$ lines of configuration commands from $53$ files, where $16$ files are real configuration files from the industry, $20$ files from Huawei configuration manual examples, and $17$ files from Nokia configuration manual examples. 
The dataset covers various settings of the routers, including basic system information, interface, route policy, filter policy,  BGP/IGP protocols, VPRN, etc.

\noindent \textbf{Metrics.}
The evaluation metrics take into account both the syntactic and view correctness, as well as the similarity to the reference configuration.
\textit{(1) Tree Match:} the matching rate on the configuration command hierarchy tree, which checks both the syntactic and view correctness.
\revised{\textit{(2) Syntax Correctness:} 
This metric is designed to measure the pure syntax correctness because view errors caused by some commands may affect the matching of subsequent syntax-correct commands on the hierarchy tree.}
\textit{(3) BLEU}~\cite{papineni2002bleu}: a widely used metric that focuses on precision, often used to evaluate the output quality of machine translation tasks. 
\textit{(4) Exact Match:} the strict matching rate, measuring the recall rate at the full command line level.
\revised{\textit{(5) Command Match:} a recall-based metric measuring whether the required commands are successfully translated, serving as a quantitative indicator of semantic correctness.}
\revised{Note that \textit{Exact Match} is intuitive but limited, as correct translations may be non-unique. \textit{Tree Match}, \textit{Syntax Correctness}, and \textit{Command Match} are more informative and task-relevant metrics.}
We use Recall Rate@Top-k to evaluate the manual retrieval module. It denotes the percentage of queries where at least one ground-truth manual appears in the top k retrieved results.

\noindent \textbf{Methods and baselines.}
We use the LLM-only translation results as the baseline for the end-to-end experiment.
\method{} is our full method, which includes the Intent-based Retrieval-Augmented Generation (IRAG) process (including intent-based manual retrieval and incremental translation), followed by syntax and semantic refinement.
We also evaluate two ablated versions of \method{}: one with only the IRAG process (IRAG-only), and one that includes both IRAG and Syntax Refinement (IRAG + Syntax).
\revised{We use GPT-4o~\cite{openai:gpt-4o}, Qwen-Max~\cite{alibaba:qwen-max} and DeepSeek-V3~\cite{liu2024deepseek} as the base models to avoid single-model bias and to ensure more convincing results.}
We also include ConfigTrans~\cite{zheng2024configtrans} as a baseline method and conduct separate comparisons within its supported scope.

\subsection{End-to-End Evaluation}

\begin{table*}[t]
    \centering
    \caption{End-to-end evaluation.}%
    \resizebox{0.9\textwidth}{!}{%
    \renewcommand{\arraystretch}{1.1}
    \begin{tabular}{llccccc}
        \toprule
        \textbf{Base Model} & \textbf{Method} & \textbf{Tree Match} & \textbf{Syntax Correctness} & \textbf{    BLEU-2    } & \textbf{Exact Match} & \textbf{Command Match} \\
        \midrule
        \multirow{4}{*}{GPT-4o} 
            & Baseline              & 0.7041 & 0.8560 & 0.6749 & 0.4980 & 0.6748 \\
            & IRAG-only              & 0.8438 & 0.9144 & 0.7727 & 0.5852 & 0.8069 \\
            & IRAG+Syntax            & 0.9219 & 0.9727 & 0.7688 & 0.6418 & 0.8155 \\
            & \method{} (Full Method)    & \textbf{0.9277} & \textbf{0.9813} & \underline{\textbf{0.7880}} & \underline{\textbf{0.6531}} & \textbf{0.8287} \\
        \hline
        \multirow{4}{*}{Qwen-Max} 
            & Baseline              & 0.6142 & 0.7466 & 0.6680 & 0.4900 & 0.6100 \\
            & IRAG-only              & 0.8586 & 0.9616 & 0.7143 & 0.5818 & 0.7911 \\
            & IRAG+Syntax            & 0.9256 & 0.9769 & 0.7367 & 0.6055 & 0.8101 \\
            & \method{} (Full Method)    & \textbf{0.9364} & \textbf{0.9781} & \textbf{0.7483} & \textbf{0.6179} & \textbf{0.8169} \\
        \hline
        \multirow{4}{*}{DeepSeek-V3} 
            & Baseline              & 0.7315 & 0.8654 & 0.7144 & 0.5959 & 0.6944 \\
            & IRAG-only              & 0.8690 & 0.9511 & 0.7284 & 0.5902 & 0.8148 \\
            & IRAG+Syntax            & 0.9727 & 0.9913 & 0.7458 & 0.6102 & 0.8256 \\
            & \method{} (Full Method)    & \underline{\textbf{0.9815}} & \underline{\textbf{0.9966}} & \textbf{0.7654} & \textbf{0.6464} & \underline{\textbf{0.8472}} \\
        \bottomrule
    \end{tabular}%
    }
    \label{tab:e2e}
\end{table*}

\noindent \textbf{Results.}
The result of the end-to-end evaluation is shown in Table~\ref{tab:e2e}.
In the table, bold font indicates the best result within each group of base models. Underlining denotes the best result across all methods in that column.
Our full method, \method{}, has demonstrated a significant improvement compared to the baseline of LLM-only translation.
With DeepSeek-V3 as the base model, the \textit{syntax correctness} rate reaches $99.66\%$ and the \textit{tree match} (view and syntax correctness) rate reaches $98.15\%$, which are $13.12\%$ and $25.00\%$ higher than baseline, respectively. 
These two metrics indicate that \method{} performs well in generating target device configurations with respect to both syntactic and view correctness.
The \textit{command match} rate also has a significant improvement, reaching $84.72\%$, which is $15.28\%$ higher than the baseline.
This metric reflects \method{}'s effectiveness in recalling the necessary commands for the target device in the translated configuration.
In addition, our method also shows its effectiveness on GPT-4o and Qwen-Max, indicating that \method{} can adapt to different base models.

\noindent \textbf{Ablation study.} 
We can see from Table~\ref{tab:e2e} that the method with only IRAG has a great improvement in translation performance compared to the baseline.
In the IRAG-only method with DeepSeek-V3 as base model, \textit{tree match} increases by $13.75\%$ compared to the baseline while \textit{command match} increases by $12.04\%$.
\revised{The results indicate that the IRAG module enhances the LLMs' ability to generate syntactically correct and semantically consistent configurations by retrieving and incorporating relevant manuals of the target device, which serve as a crucial source of domain-specific knowledge for accurate translation.}
Table~\ref{tab:e2e} also shows that our modules for syntax and semantic refinement have a significant effect on improving the translation quality.
With syntax refinement, the \textit{tree match} rate increases from $86.90\%$ to $97.27\%$.
This indicates that syntax checks and refinements based on command hierarchy trees and LLM can significantly improve syntax and view correctness in the incremental translation process.
The semantic verification and refinement module improves command match and exact match by 2.16\% and 3.62\% respectively (from IRAG+Syntax to \method{}). 
It also plays a key role in generating translation quality reports tailored for human engineers, the accuracy of which is evaluated in \Section~\ref{sec:report}.

\begin{table}[t]
    \centering
    \caption{\revised{Comparison with ConfigTrans.}}
    \label{tab:configtrans}
    \resizebox{0.95\linewidth}{!}{  
        \begin{tabular}{lcc}
        \toprule
        \textbf{Method} & \textbf{Acc. (w/ param.)} & \textbf{Acc. (w/o param.)} \\
        \midrule
        \method{} (with GPT-4o) & 0.8615 & 0.8957 \\
        ConfigTrans & 0.8247 & 0.7550 \\
        \bottomrule
        \end{tabular}
    }
    \vspace{-1em}
\end{table}

\noindent \revised{\textbf{Comparison with ConfigTrans.}  
We conduct experiments on the dataset used in ConfigTrans~\cite{zheng2024configtrans}, which includes BGP and OSPF commands. 
As shown in Table~\ref{tab:configtrans}, \method{} outperforms ConfigTrans in translation accuracy for both parameterized and non-parameterized commands.
More importantly, unlike ConfigTrans, \method{} does not rely on mode-specific heuristic algorithms and pre-defined parameter correspondence tables, enabling better generalization and lower migration overhead when adapting to new device vendors (Section~\ref{sec:overhead}).}

\noindent \revised{\textbf{Performance on smaller LLMs.}
We also evaluate Llama3.1-8B~\cite{dubey2024llama} and Qwen3-8B~\cite{yang2025qwen3}, and observe two key limitations.
(1) They struggle to produce well-formatted outputs, often fail to generate valid JSON during intent extraction and configuration splitting. %
(2) They lack sufficient understanding of configuration semantics and vendor-specific knowledge, resulting in lower translation quality. 
For successful cases, Llama3.1-8B achieves 55.96\% syntax accuracy and 41.91\% command match rate, while Qwen3-8B reaches 73.76\% and 55.18\%. 
The results show that small-scale 8B LLMs remain inadequate for configuration translation.
}

\noindent \textbf{Case study.}
In Appendix~\ref{sec:case_study}, we use a concrete example to show the step-by-step translation process of INTA.

\subsection{Intent-Based Manual Retrieval Module}\label{sec:ra-module}

\begin{figure}[t]
    \centering
    \includegraphics[width=0.48\textwidth]{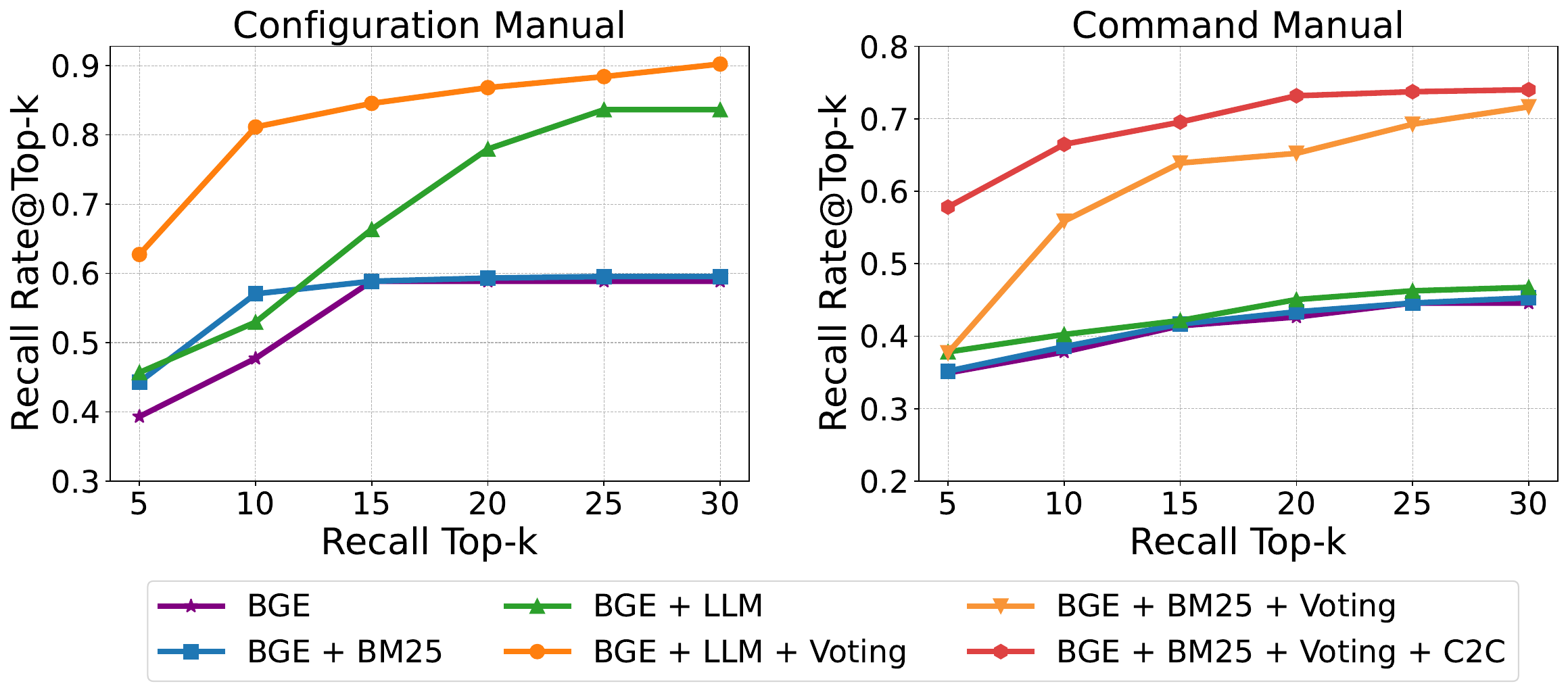}
    \caption{Ablation study of the intent-based retrieval module.}
    \label{fig:recall-ablation}
    \vspace{-1em}
\end{figure}

To evaluate the recall rate of the intent-based manual retrieval module, we manually annotate 409 mappings from source device configurations to target device configuration/command manuals, which serve as the test set for module evaluation. 
We use Qwen-Max as the default LLM base model. %

We conduct an ablation study to verify the effectiveness of each module in the manual retrieval process.
The results are shown in Fig.~\ref{fig:recall-ablation}.
The baseline method is \textit{BGE}, which uses only the BGE embedding model for retrieval.
\textit{LLM} and \textit{BM25} serve as preliminary filtering strategies.
\textit{Voting} denotes a voting mechanism that combines the retrieval results of multiple intents.
\textit{C2C} refers to retrieving related command manuals based on the retrieved configuration manuals. %

For configuration manual retrieval, using \textit{LLM} as a preliminary filter significantly improves the ``tail-end'' performance (Top-20 to Top-30), while \textit{BM25} has almost no effect on improving the recall rate.
Building on the \textit{LLM} filter, the \textit{Voting} mechanism further improves the ``head-end'' recall rate (Top-5 to Top-20), indicating that aggregating retrieval results based on fine-grained intent interpretations effectively boosts the retrieval of relevant manuals.

For command manual retrieval, neither \textit{BM25} nor \textit{LLM} improves retrieval accuracy effectively, suggesting that the features of command manuals are more difficult to capture.
Nevertheless, we opt for the more efficient and lightweight \textit{BM25} to reduce the candidate manual set and accelerate the subsequent dense retrieval process.
The \textit{Voting} mechanism significantly enhances overall performance, particularly improving the recall rate in the ``tail-end''.
Meanwhile, the \textit{C2C} mechanism further boosts the ``head-end'' recall rate, indicating that the configuration information contained in accurately retrieved configuration manuals is highly relevant and critical for command manual retrieval.

\subsection{Deep Dive}

\begin{figure}[t]
    \centering
    \vspace{0.5em}
    \includegraphics[width=0.48\textwidth]{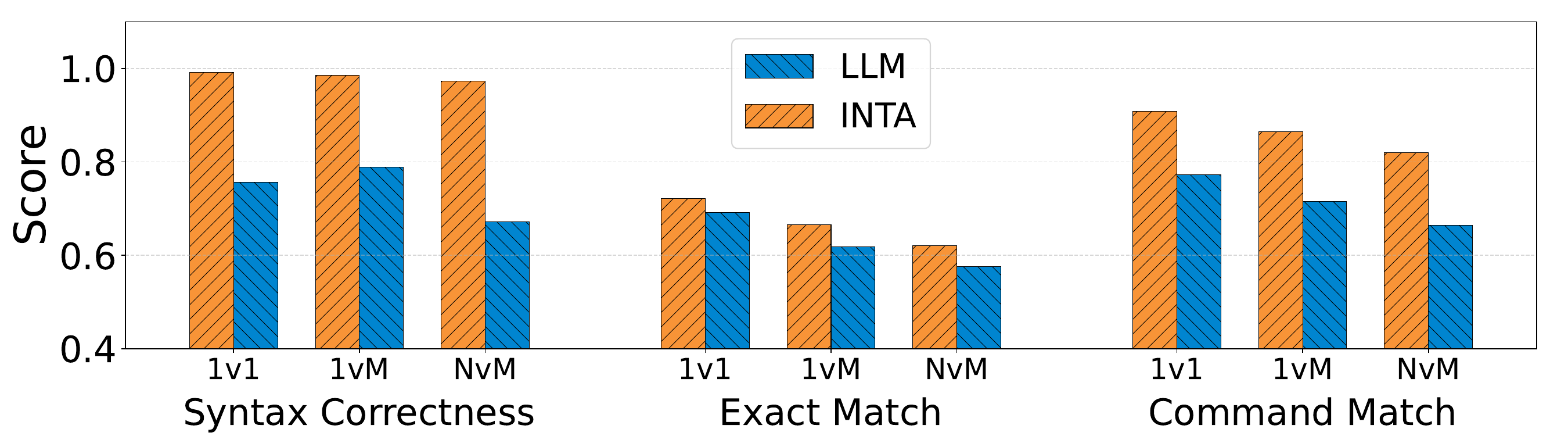}
    \caption{Deep dive into varying levels of translation difficulty.}
    \label{fig:deep-dive}
\end{figure}

\noindent \textbf{Different difficulty levels.}
\revisednochange{
To further evaluate the performance of \method{}  under varying levels of translation difficulty, we divide the dataset into three types:
(1) one source command maps one target command (1v1); 
(2) one source command maps $M$ target commands (1vM);
(3) $N$ source commands map $M$ target commands (NvM).
The 1vM category also includes cases where $M$ source commands map to one target command.
Fig.~\ref{fig:deep-dive} shows the results using DeepSeek-V3 as the base model.
\method{} consistently outperforms LLM-only translation across all types.
Notably, on the most challenging NvM subset, it achieves the largest gain in syntax correctness.
These results underscore \method{}’s effectiveness, particularly in handling complex configuration translation scenarios.}

\noindent \revised{\textbf{Different input lengths.}
Additional experiments demonstrate that \method{} maintains stable performance across different input configuration lengths within a certain range, with detailed results provided in Appendix~\ref{sec:length}.}

\noindent \revised{\textbf{Configuration name preservation.}
We further analyzed \method{}'s translation results 
and observed that 96.49\% of configuration names (e.g., interface or route policy) remained consistent with the source when both source and target configurations required explicit naming.}

\subsection{Human Evaluation for Translation Report}\label{sec:report}

\begin{table}[t]
    \centering
    \caption{Quality of translation reports.}
    \label{tab:human-eval}
    \begin{tabular}{lccccc}
    \toprule
    \textbf{Model} & \textbf{TN} & \textbf{TP} & \textbf{FN} & \textbf{FP} & \textbf{Accuracy}\\
    \midrule
    GPT-4o & 73 & 10 & 0 & 6 & 93.26\%\\
    Qwen-Max & 64 & 9 & 0 & 10 & 87.95\%\\
    DeepSeek-V3 & 87 & 10 & 0 & 8 & 92.38\% \\
    \bottomrule
    \end{tabular}
    \vspace{-1em}
\end{table}

To assess the quality of the translation reports, we conduct a manual evaluation with industry experts using a Huawei NE40E router in the GNS3 emulator.
We randomly select 12 reports and evaluate all report units within them.
The number of report units varies by model.
The evaluation metrics include \textit{True Negative (TN)}, \textit{True Positive (TP)}, \textit{False Negative (FN)}, \textit{False Positive (FP)}, and \textit{Accuracy}.
Here, we treat \textit{incorrect configurations} as \textit{positive} cases.
The evaluation results are shown in Table~\ref{tab:human-eval}.
It shows that the LLMs provide translation reports with relatively high accuracy.
They do not miss incorrect configurations (i.e., zero false negatives), but sometimes misclassify correct configurations as incorrect ones (i.e., false positives).
This is mainly because LLMs are too rigorous in capturing the literal differences between configurations, leading to misclassifying some correct configurations as incorrect ones.
Nevertheless, thanks to the high recall rate for incorrect configurations, the reports still offer practical guidance for operational use.

\subsection{Exploring Generalization Capability}\label{sec:generality}

\begin{table}[t]
    \centering
    \caption{Cisco-to-Huawei switch scenario evaluation.}
    \label{tab:cisco-huawei}
    \resizebox{\linewidth}{!}{  %
        \begin{tabular}{lcccc}
        \toprule
        \textbf{Method} & \textbf{Syntax Correctness} & \textbf{Exact Match} & \textbf{Command Match} \\
        \midrule
        LLM (DeepSeek-V3) & 0.7188 & 0.3742 & 0.6950 \\
        \method{} (DeepSeek-V3)   & 0.9650 & 0.6431 & 0.7884 \\
        \bottomrule
        \end{tabular}
    }
    \vspace{-1em}
\end{table}

To validate the effectiveness of our approach in other scenarios, we conduct an evaluation on translating configurations from Cisco to Huawei switches.
We select 146 configuration cases from the Cisco configuration manual as the test set. The evaluation results are shown in Table~\ref{tab:cisco-huawei}.
In this scenario, \method{} also demonstrates significant improvements over the LLM baseline.
The results validate \method{}'s generalization capability across different scenarios.

\subsection{\revised{Overhead}}\label{sec:overhead}
\noindent \revised{\textbf{Runtime overhead.}}
\revised{
We define \textit{runtime overhead} as the average time and monetary cost \textit{per line} (translating one line of source device configuration).
Using Qwen-Max as the base model, the full \method{} pipeline takes an average of 16.12 seconds per line on an NVIDIA A100 GPU, and 30.93 seconds on an NVIDIA RTX 3090 GPU.
As for LLM cost, it consumes an average of 3536 prompt tokens and 276 completion tokens per line. 
This corresponds to approximately \$0.0015 per line with Qwen-Max, or \$0.0116 with GPT-4o.
Based on our practical experience and industry data, this process typically takes hours per line with high labor costs.
\method{} achieves substantially lower time and monetary overhead.
}

\noindent \revised{\textbf{Migration overhead.}
We define \textit{migration overhead} as the total effort required to support a new device vendor, measured by (1) lines of code (LoC) for system extension and (2) estimated human effort in person-days.
The estimated migration overheads are as follows:
\method{} requires $\sim$100 LoC and 2 person-days per new device, involving only manual scraping of device manuals and minor parser modifications, without the need for predefined mapping rules.
ConfigTrans incurs substantially higher overhead, requiring a $\sim$500 LoC and 14 person-days per configuration mode due to the need for heuristic algorithm design and command parameter mapping table construction.
NAssim, although not an end-to-end system, has a similar migration overhead to \method{}.} %

\section{Discussion}

\subsection{Practical Application Value}

\method{} significantly reduces cost and time compared to traditional methods that rely heavily on manual effort during vendor migration.
It delivers reliable accuracy in generating translated configurations, with no false positives observed in translation reports across the evaluated dataset.
This enables users to focus solely on flagged errors, substantially reducing the workload for human experts.
\revised{Even when occasional false positives arise, the translated configurations and accompanying reports still provide valuable guidance, requiring far less effort than translating from scratch.}
While not yet fully autonomous, \method{} already serves as a practical assistant by offering relatively reliable translations to assist human experts.

\subsection{Limitations and Future Work} \label{sec:limitation_future}

Although \method{} performs well in cross-vendor configuration translation, it still has limitations and suggests future work.

\noindent \revised{\textbf{Unified context tracking:}
\method{} currently employs incremental translation to manage dependencies across configuration fragments. Future work includes developing a more robust and unified context tracking mechanism.}

\noindent \revised{\textbf{Hybrid verification:}
The \method{} framework is currently using a syntax tree and LLM for verification and refinement. Future work includes formal and simulation-based verification.}

\noindent \textbf{Real device interaction:} 
Future improvements also include incorporating real-time device status (e.g., port status) to enhance the practicality and accuracy of translation.

\noindent \revised{\textbf{Network-wide scenarios:} 
\method{} currently focuses on translating configurations for individual devices. 
Future work includes incorporating topology-aware context and modeling inter-device dependencies to support network-wide scenarios.}

\noindent \textbf{SDN/NFV integration:}
Future work also includes extending \method{} to support dynamic, virtualized, and service-oriented SDN/NFV scenarios for more efficient network management.

\section{Related Work} \label{sec:related_work}
\revised{
NAssim~\cite{chen2022software} generates and validates configuration models from device manuals to integrate legacy devices into SDN networks.}
\revised{
ConfigTrans~\cite{zheng2024configtrans} combines constraint solving and LLMs for configuration translation.}
\revised{
Nissist~\cite{an2024nissist} uses LLMs to clarify intent and extracts information from knowledge bases for network troubleshooting.}
\revised{
Lumi~\cite{jacobs2021hey} translates human intents into an abstraction layer \textit{Nile} via machine learning, but can't generate low-level device-specific configuration commands.}
\revised{Nissist and Lumi primarily focus on capturing and analyzing human intent, while INTA introduces a novel approach by using intent to bridge heterogeneous configuration models.}
\revised{Batfish~\cite{fogel2015general} uses a rule-based intermediate vendor-neutral representation for network configuration analysis, but it suffers from limited coverage and high migration overhead.}
\revised{
CEGS~\cite{liu2025cegs} automates network configuration synthesis using graph neural networks (GNNs) and LLMs, but its capability and coverage rely on high-quality configuration examples.}    %
Verified Prompt Programming~\cite{mondal2023llms} combines GPT-4 with verifiers to generate correct router configurations. 

\section{Conclusion}

Traditional network configuration translation methods are labor-intensive due to diverse configuration models.
We propose \method{}, an LLM-driven intent-based framework with four key modules to address this challenge.
Experiments on industry datasets show that \method{} outperforms existing methods and exhibits generalization across various network scenarios, highlighting its practical value for real-world applications.

\newpage

\bibliographystyle{plain}

\bibliography{references.bib}

\newpage

\appendices

\section{Configuration Parser Details}\label{appdx:parser-details}
\revised{We use the configuration parser in \method{} for two purposes:
(1) to extract the exact manual pages corresponding to the source configurations, which facilitates configuration division and intent extraction;
(2) to verify the syntax correctness of the translated configurations and guide syntax refinement during the incremental translation process.}
The parser consists of two interdependent core modules: the command syntax parser and the command hierarchy parser.
We construct the parser using an approach similar to that of NAssim~\cite{chen2022software}.

\begin{figure}[t]
    \centering
    \includegraphics[width=0.45\textwidth]{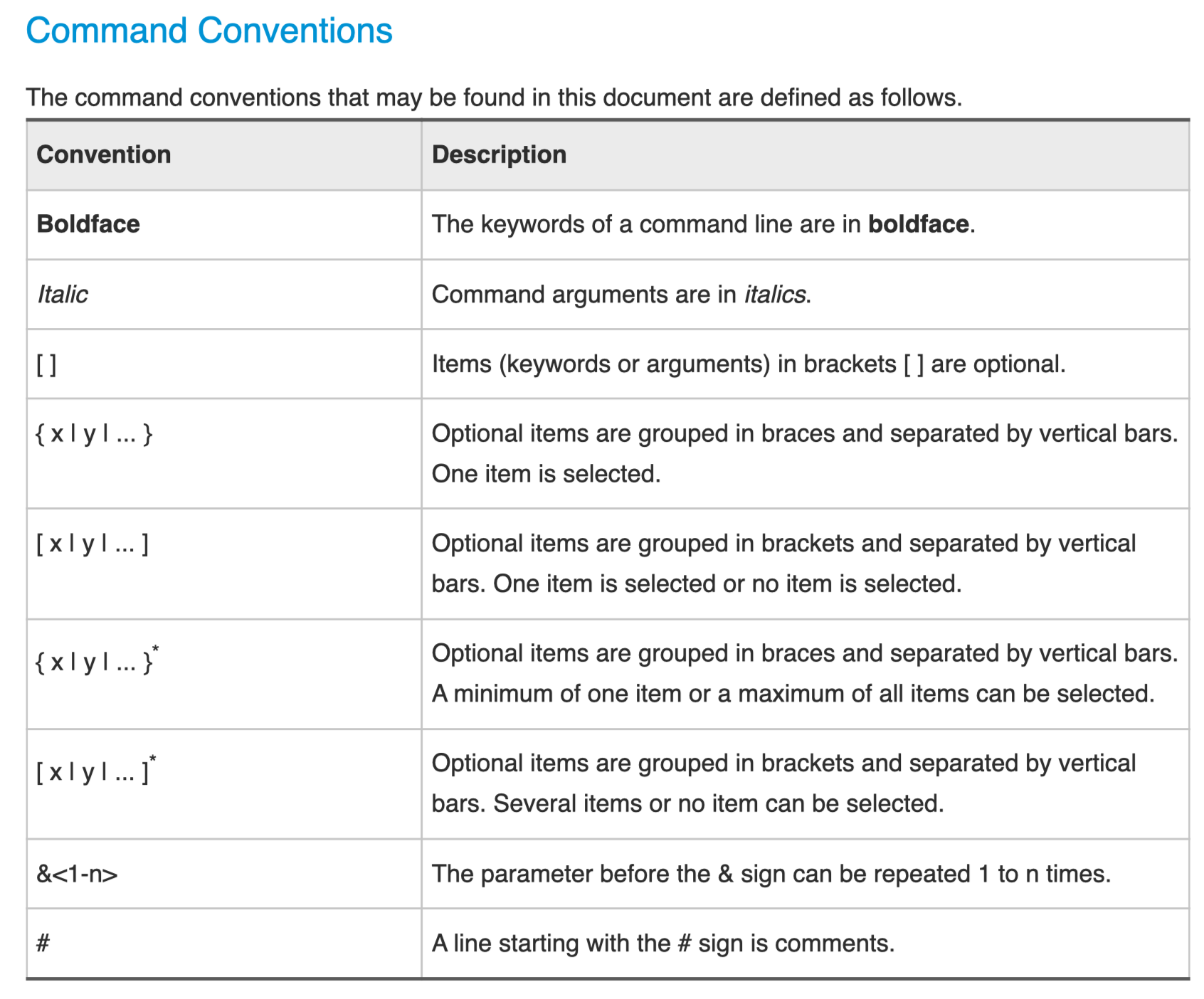}
    \caption{Command conventions of Huawei command manuals.}
    \label{fig:cmd-convention}
\end{figure}

\noindent \textbf{Command syntax parser.}
The command syntax parser constructs a command graph based on the command manual, which serves as a matching template.
The construction of the parser relies on the command template conventions described in the command manual, as illustrated in Fig.~\ref{fig:cmd-convention}.
These command graphs enable line-by-line parsing of commands and extraction of their syntax trees.
The command graph includes three types of non-leaf nodes: \verb|seq| (sequential nodes), \verb|req_selector| (corresponding to \verb!{x|y|...}!), and \verb|opt_selector| (corresponding to \verb![x|y|...]!).
Other conventions, such as \verb|*| and \verb|&| sign, are implemented as features of the nodes.
It also contains two types of leaf nodes: keyword nodes and parameter nodes.
In addition, there are two functional node types: the \verb|end| node (indicating the end of a command) and the \verb|pass| node (serving as a placeholder in \verb|opt_selector|).
For a given command template (e.g., the format shown in Fig.~\ref{fig:command_manual_example}), we traverse the command format specification from left to right to construct the command graph.
The command graph is built using a recursive structure, starting with an outermost \verb|seq| node. When encountering an optional item, a corresponding \verb|req_selector| or \verb|opt_selector| node is created.
The constructed command graph is then used to match actual configuration commands, associating elements in the configuration (such as keywords and parameters) with corresponding elements in the command graph.

\noindent \textbf{Command hierarchy parser.}
Based on the command syntax parser, we build a command hierarchy parser to analyze the hierarchical view structure of configuration files.
The command hierarchy parser relies on the device's command hierarchy file, namely the VDM in NAssim~\cite{chen2022software}.
A VDM file is a JSON file that specifies, for each command, its type, CLI definition, associated view, and all subcommands (children) available within the view it enters.
For a complete configuration file, starting from the root node of the VDM (e.g., \verb|system-view| command in Huawei devices' system view), the hierarchy parser uses the command syntax parser to parse the configuration file line by line, match commands in the VDM, and record the corresponding view hierarchy transitions.
The parsed output not only contains the syntax matching results for each command but also includes the hierarchical view context to which each command belongs.

\noindent \revised{\textbf{Example of adapting to a new vendor.}}
\revised{To illustrate the migration effort required when adapting the parser to a new vendor, we provide a concrete example using Huawei NE40E.}

\revised{The \textit{Command Syntax Parser} is decoupled from vendor-specific details and only relies on a unified command template convention (Fig.~\ref{fig:cmd-convention}). Therefore, as long as the target vendor’s documentation can be transformed into this convention, the syntax parser remains fully reusable without any modification.}

\revised{The \textit{Command Hierarchy Parser} is vendor-specific and reflects differences in configuration models. Taking Huawei as an example, the main modification lies in supporting a large number of interface views. In the Huawei NE40E router, simply recognizing the \texttt{interface} command does not lead to a single fixed view because there are 109 possible interface view types. Therefore, the parser must recognize the exact interface type (e.g., GE, Loopback, VLANIF) to determine the correct sub-view to enter. Moreover, the parser must also distinguish between sub-interface views and other variants. As a result, the main adaptation effort for Huawei NE40E lies in handling these diverse interface views.}

\section{Manual Examples}\label{sec:manual_examples}

We use two types of manuals in our work: \textit{Command Manuals} and \textit{Configuration Manuals}. We use Huawei NE40E Router manuals as examples to illustrate them. %

\begin{figure}[htbp]
    \centering
    \includegraphics[width=0.45\textwidth]{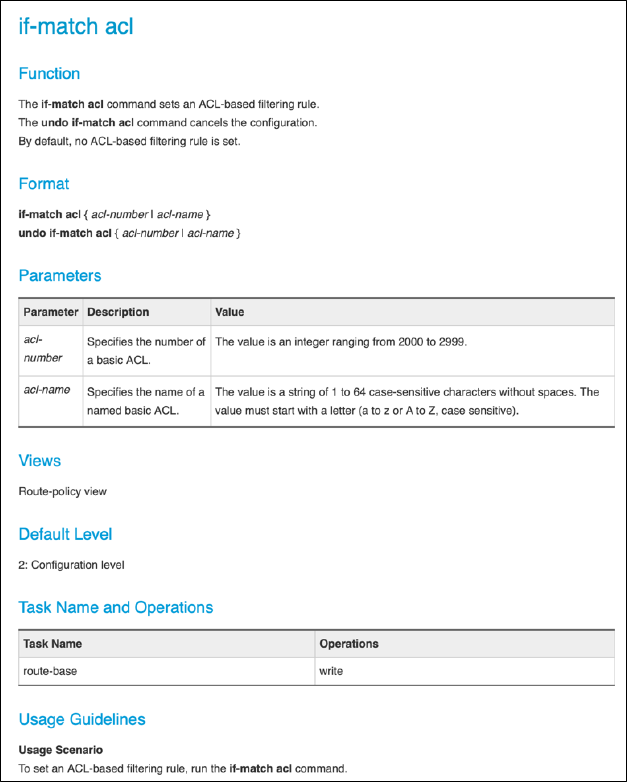}
    \caption{Command manual example.}
    \label{fig:command_manual_example}
\end{figure}

\noindent \textbf{Command manuals.} Contains syntax definitions and functional descriptions of all commands of the device, mainly used for constructing the configuration syntax tree in the parser.
Command manuals are also retrieved to enhance the translation process.
A command manual example is shown in Fig.~\ref{fig:command_manual_example}.

\begin{figure}[htbp]
    \centering
    \includegraphics[width=0.45\textwidth]{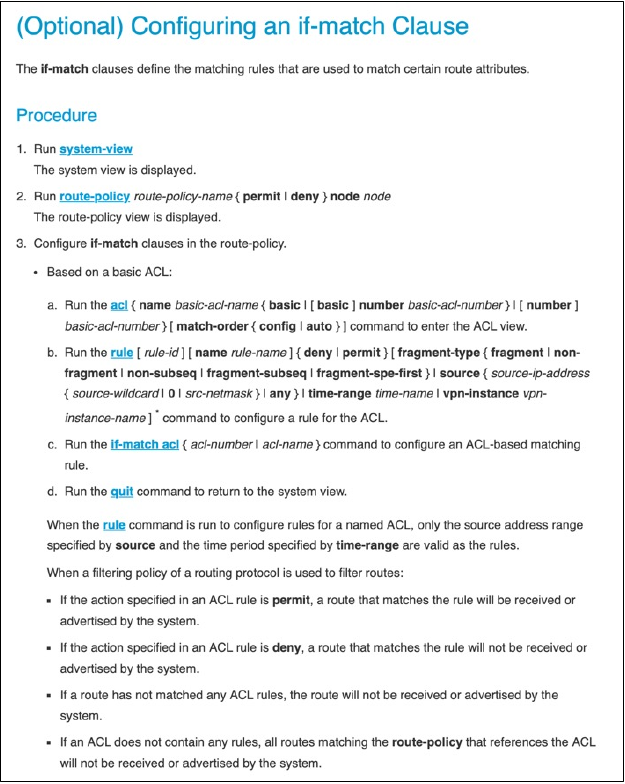}
    \caption{Configuration manual example.}
    \label{fig:config_manual_example}
\end{figure}

\noindent \textbf{Configuration manuals.} Contains configuration steps (command sequence) required to implement a certain function. 
We use extracted intents to retrieve the corresponding configuration manuals, which are used to generate the configuration.
A configuration manual example is shown in Fig.~\ref{fig:config_manual_example}.

\section{Details and Prompt Templates for Each Module}\label{sec:prompts}

In this appendix, we provide some details and prompt templates used in each module of \method{}.

\noindent \textbf{System prompt.}
The unified system prompt used in \method{} is: \textit{You are a very helpful assistant with great expertise in network operations and maintenance.}

\begin{figure}[t]
    \centering
    \includegraphics[width=0.45\textwidth]{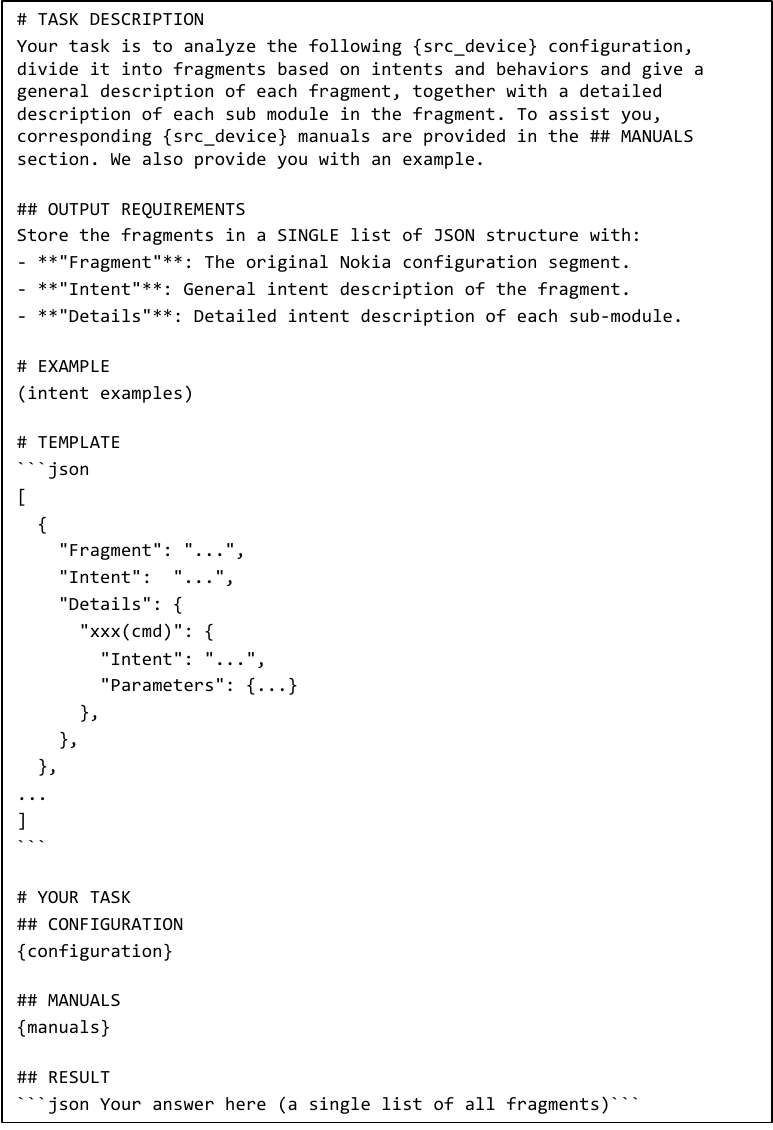}
    \caption{Example prompt template for intent extraction.} 
    \label{fig:IRAG-ICL}
\end{figure}

\begin{figure}[t]
    \centering
    \includegraphics[width=0.45\textwidth]{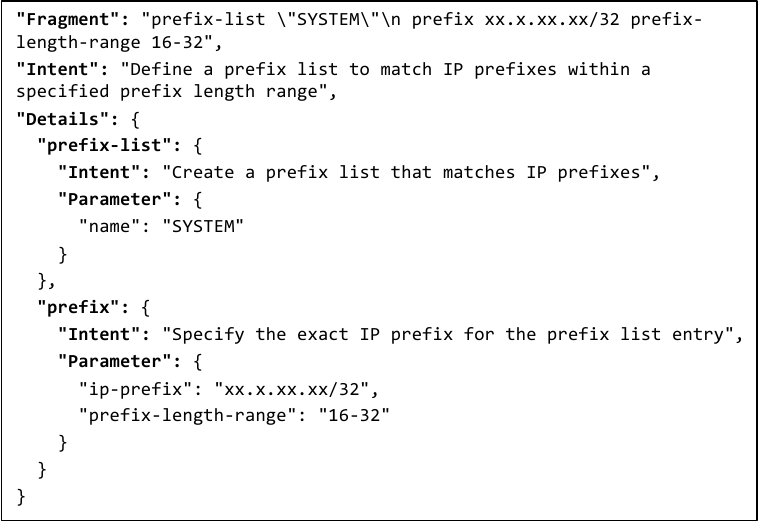}
    \caption{Example fragment of configuration division and intent extraction result.}
    \label{fig:conf-div}
\end{figure}

\noindent \textbf{Configuration division and intent extraction.}
The prompt template for intent extraction is shown in Fig.~\ref{fig:IRAG-ICL}.
We provide LLM with a JSON format instruction, together with an output template indicating the expected output format.
An example fragment of configuration division and intent extraction result is shown in Fig.~\ref{fig:conf-div}.

\begin{figure}[htbp]
    \centering
    \includegraphics[width=0.45\textwidth]{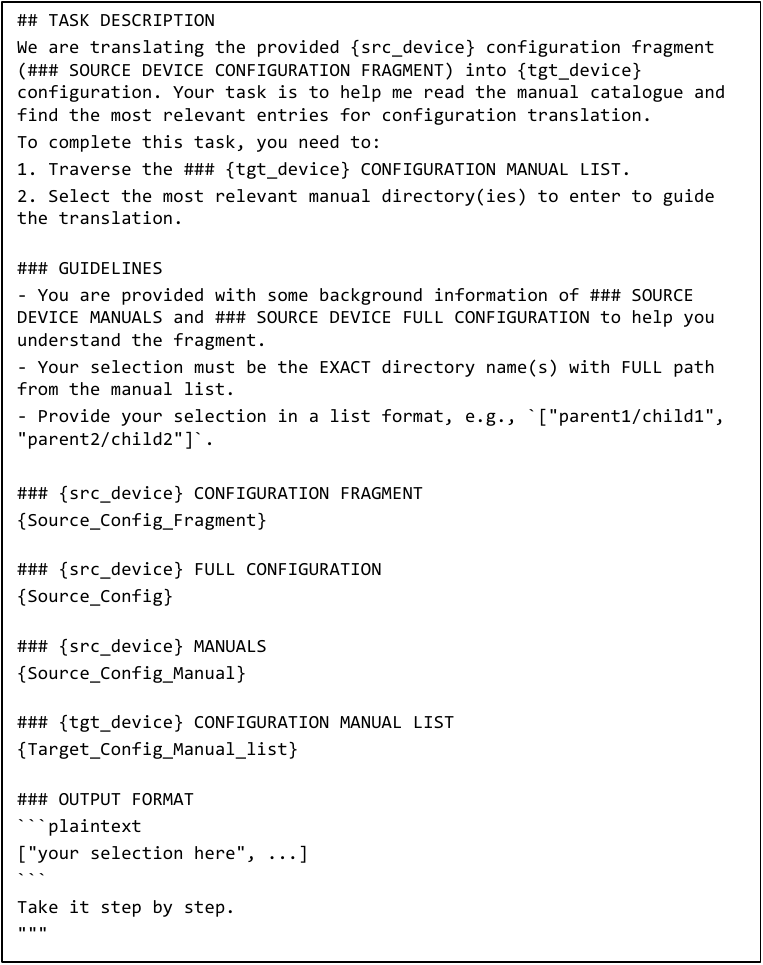}
    \caption{Example prompt template for LLM corpus filter.}
    \label{fig:IRAG-LLM-Filter}
\end{figure}

\begin{figure}[htbp]
    \centering
    \includegraphics[width=0.45\textwidth]{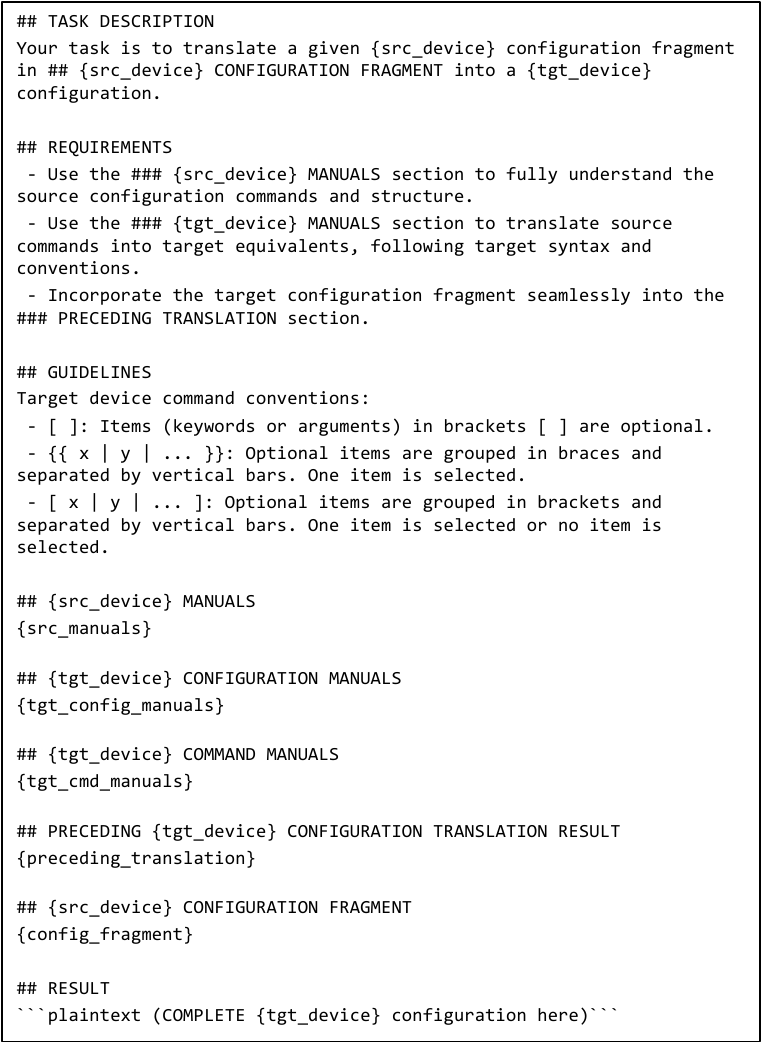}
    \caption{Example prompt template for incremental translation.}
    \label{fig:IRAG-Translation}
\end{figure}

\begin{figure}[htbp]
    \centering
    \includegraphics[width=0.45\textwidth]{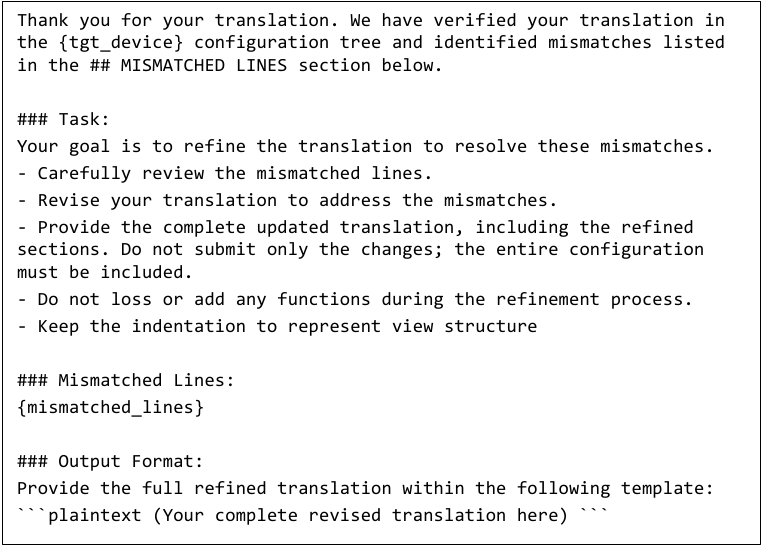}
    \caption{Example prompt template for syntax refinement.}
    \label{fig:ver-syntax}
\end{figure}

\noindent \textbf{LLM corpus filter.}
The prompt template for LLM corpus filter is shown in Fig.~\ref{fig:IRAG-LLM-Filter}. 
The output of the LLM corpus filter is a list of command/configuration manual paths for further retrieval.

\noindent \textbf{Incremental translation.}
The prompt template of incremental translation and syntax refinement is shown in Fig.~\ref{fig:IRAG-Translation} and Fig.~\ref{fig:ver-syntax}, respectively.

\begin{figure}[t]
    \centering
    \includegraphics[width=0.45\textwidth]{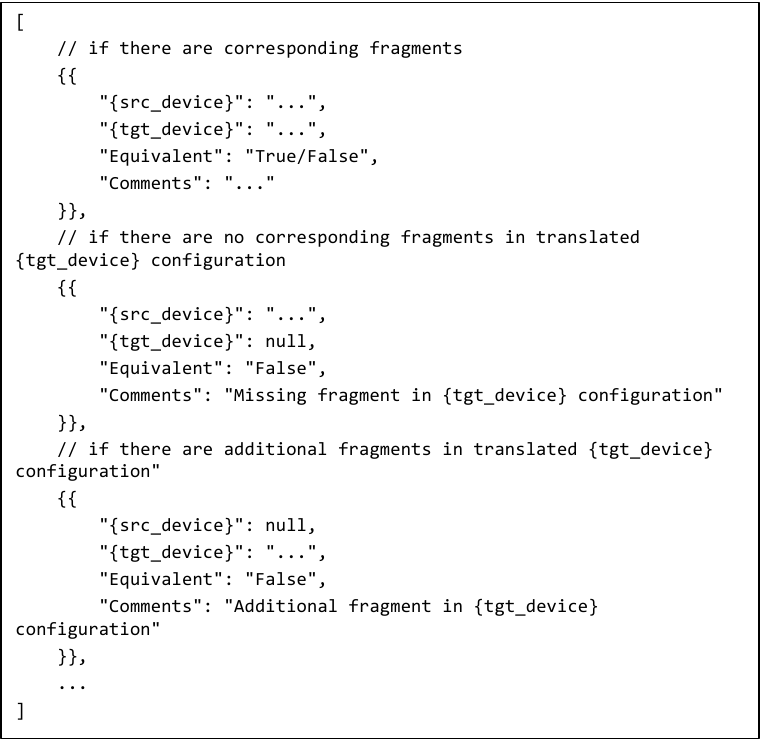}
    \caption{Report template for semantic verification.}
    \label{fig:ver-semantic-report-template}
\end{figure}

\begin{figure}[t]
    \centering
    \includegraphics[width=0.45\textwidth]{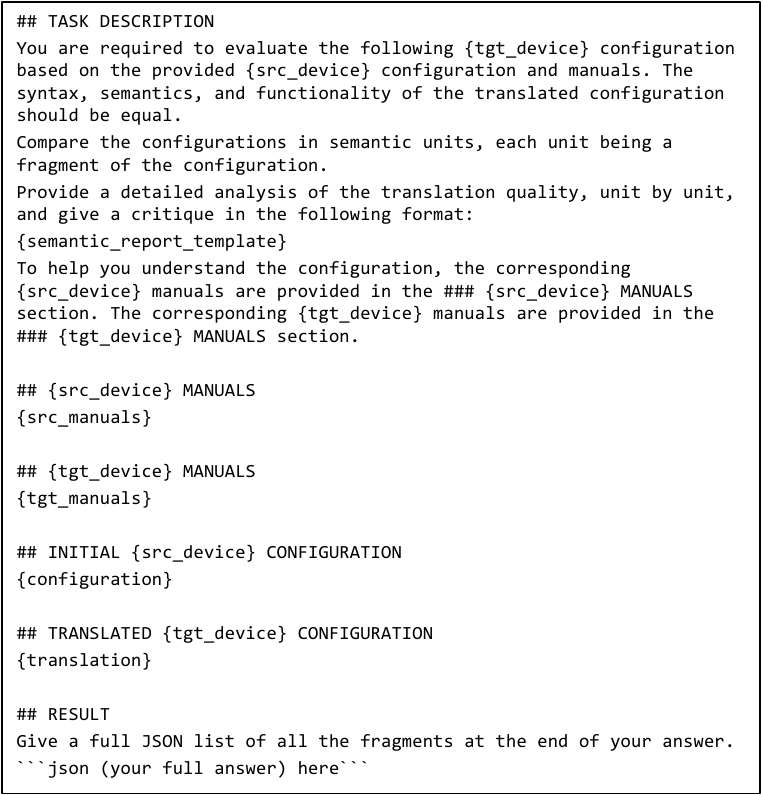}
    \caption{Example prompt template for semantic verification.}
    \label{fig:ver-semantic-prompt-template}
\end{figure}

\begin{figure}[t]
    \centering
    \includegraphics[width=0.45\textwidth]{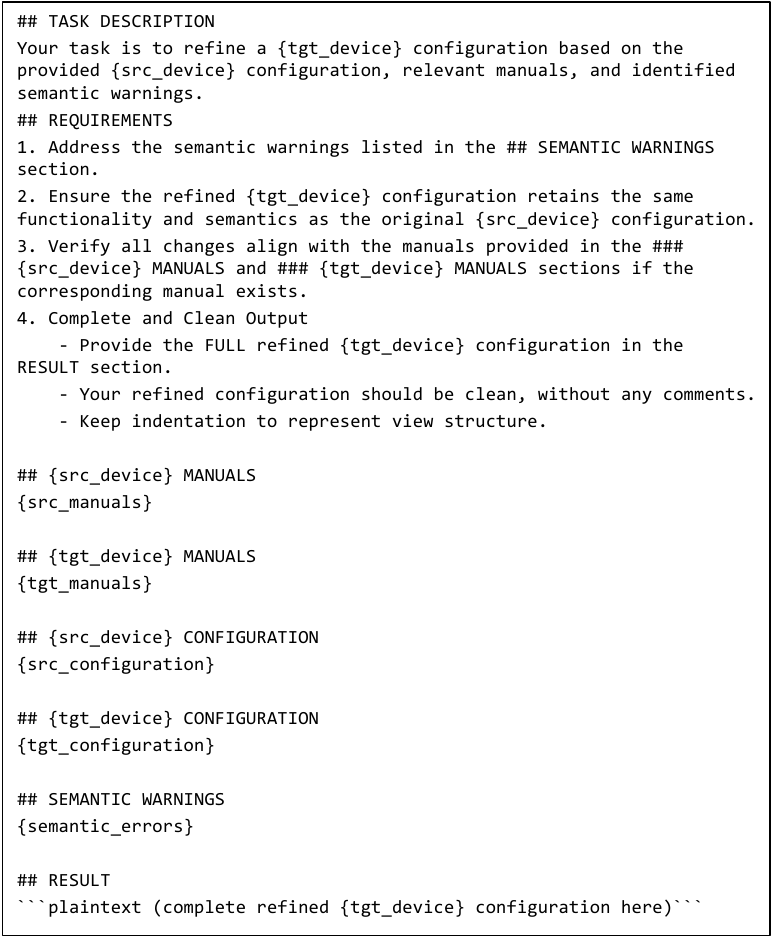}
    \caption{Example prompt template for semantic refinement.}
    \label{fig:ver-semantic-refinement}
\end{figure}

\noindent \textbf{Verification.}
Syntax verification uses the configuration tree (Appendix~\ref{appdx:parser-details}) to analyze the syntax errors.
Semantic verification and correction are both implemented with LLMs.
The prompt template for semantic verification and refinement is shown in Fig.~\ref{fig:ver-semantic-prompt-template} and Fig.~\ref{fig:ver-semantic-refinement}, respectively.
Fig.~\ref{fig:ver-semantic-report-template} shows the semantic report template used in Fig.~\ref{fig:ver-semantic-prompt-template}.

\begin{figure*}[t]
    \centering
    \includegraphics[width=0.98\textwidth]{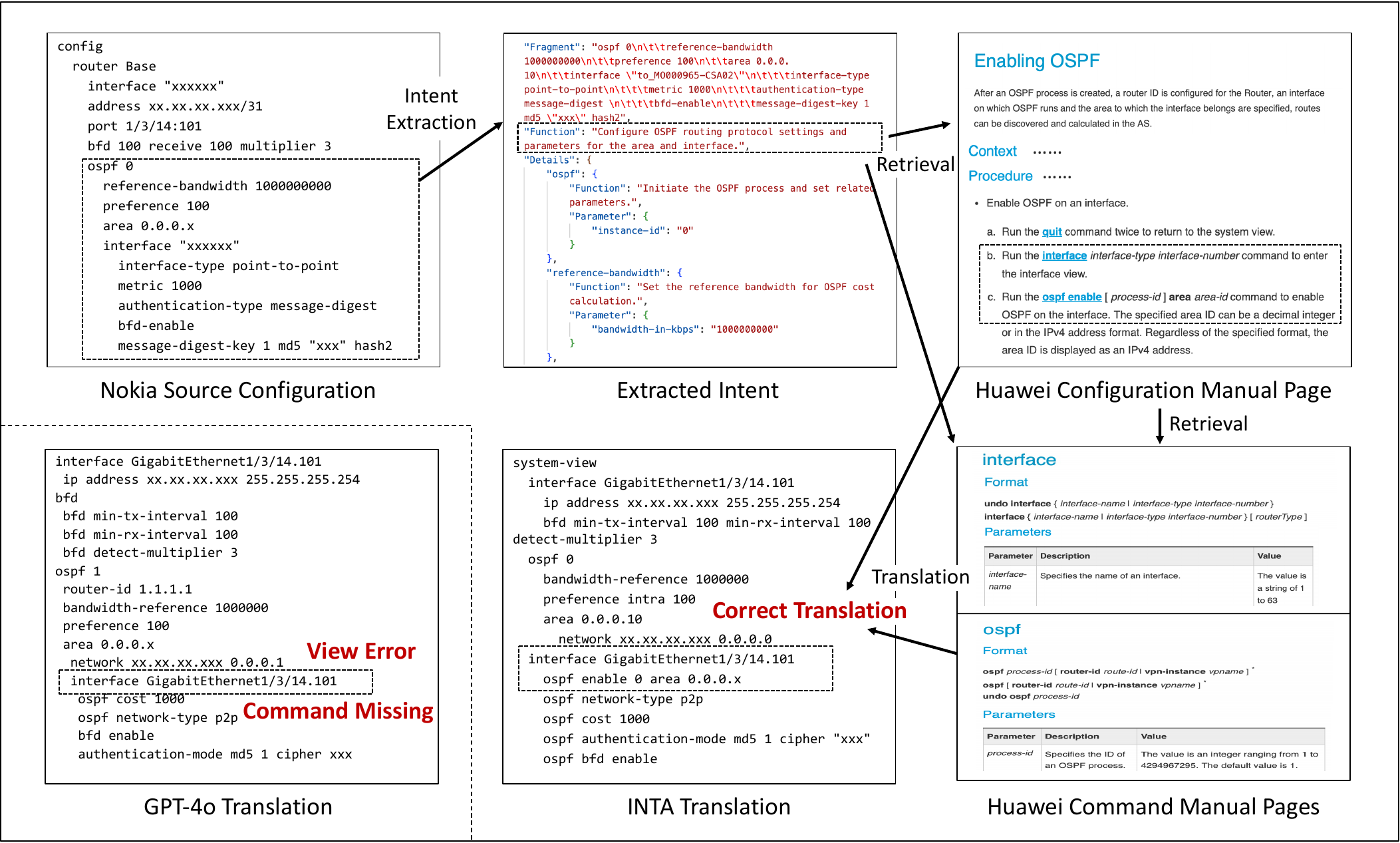}
    \caption{A case study of the configuration translation process with \method{}.}
    \label{fig:case-study} 
\end{figure*} 

\section{Case Study}\label{sec:case_study}

We provide a case study of the configuration translation process with \method{} in Fig.~\ref{fig:case-study}.
To translate the fragment in NOKIA Source Configuration, the intent extraction module extracts the intent from the fragment.
Then, the target manual retrieval module uses the extracted intents to retrieve the corresponding configuration and command manual pages.
The incremental translation module uses the retrieved manual pages and the source configuration fragment to generate the translated configuration fragment.
The target command \verb|ospf enable| is translated successfully with the correct view (interface view). This is because both the command and the view information are included in the manual page.
In contrast, the translation result of GPT-4o not only misses this command but also provides incorrect view information.

\section{\revisednochange{Parameter and Model Selection for Intent-Based Manual Retrieval Module}}\label{sec:parameter}

\begin{figure}[t]
    \centering
    \includegraphics[width=0.48\textwidth]{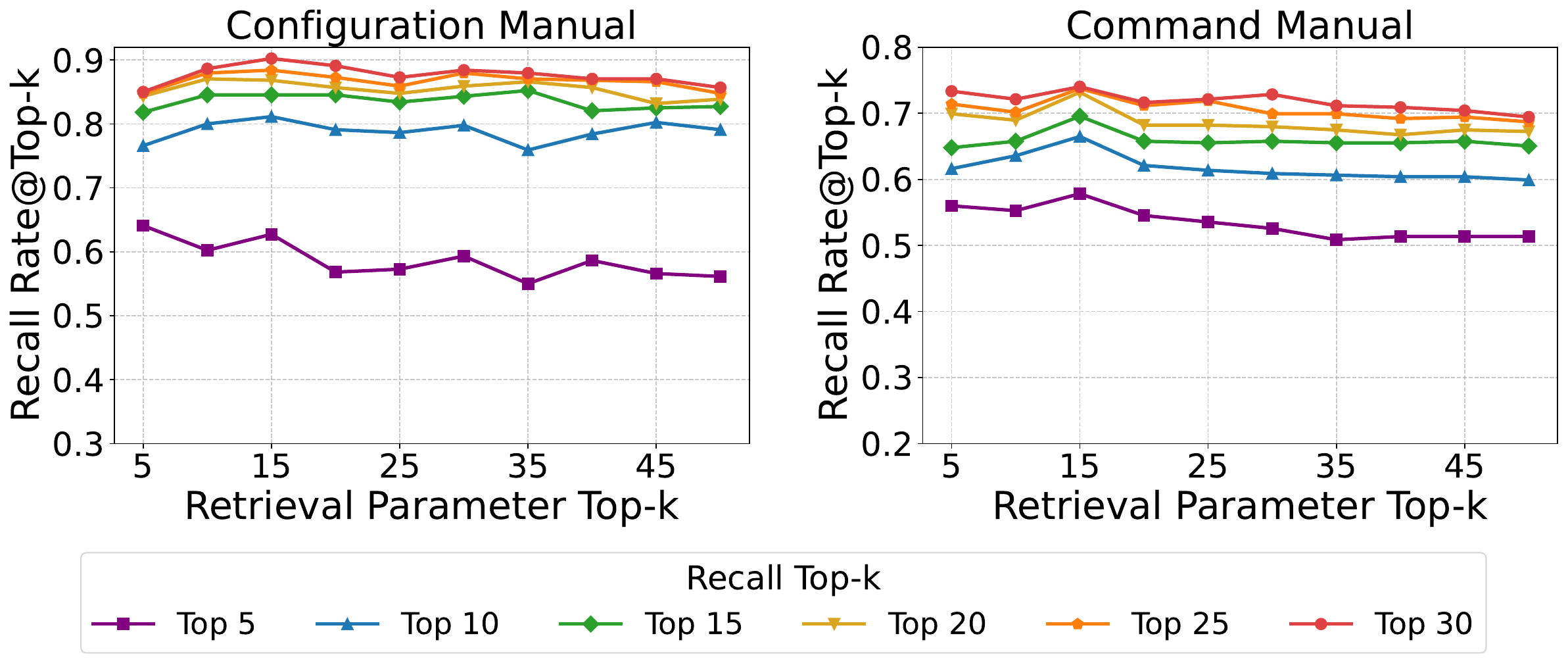}
    \caption{Relationship between single entry retrieval parameter Top-k and overall Recall Rate@top-k.}
    \label{fig:ir-topk}
\end{figure}

\noindent \textbf{Parameter selections.}
In the intent-based manual retrieval module, the number of manuals retrieved for a single entry of intent description, denoted as top‑$k$, is a key parameter. We analyze the effect of this parameter on the overall recall rate, as shown in Fig.~\ref{fig:ir-topk}.
In the figure, the choice of top‑$k$ has some impact on the overall Recall Rate@Top-k.
The best performance is achieved when $k$=15. %
In the actual translation process, considering the trade-off between LLM usage cost (input context length) and accuracy, we select the top 20 final retrieved manuals as input to the incremental translation module as their recall rate is close to that of the top 30.

\begin{table}[t] 
    \centering
    \caption{Model comparison on Configuration and Command Manual Recall Rate@Top-30.}
    \label{tab:model_performance}
    \resizebox{0.9\linewidth}{!}{  
        \begin{tabular}{lccc}
        \toprule
        \textbf{Metric} & \textbf{Qwen-Max} & \textbf{GPT-4o} & \textbf{DeepSeek-V3} \\
        \midrule
        Configuration Manual & 0.9023 & 0.8617 & 0.8662 \\
        Command Manual & 0.7402 & 0.6578 & 0.7470 \\
        \bottomrule
        \end{tabular}
    }
\end{table}

\noindent \textbf{Performance of different LLMs and embedding models.}
We compare the performance of different LLMs in terms of Recall Rate@Top-30 on configuration and command manuals, as shown in Table~\ref{tab:model_performance}. 
All three models perform well across both types of manuals, with minor variations. 
We adopt the BGE-M3 model (with 568M parameters) as our default embedding model, striking a balance between efficiency and performance.
For comparison, we also evaluated gte-Qwen2-1.5B-Instruct (with 1.5B parameters).
However, the latter achieved Recall Rate@Top-30 of $79.63\%$ and $71.76\%$ on the configuration and command manuals, respectively, both slightly lower than those of the BGE model.
This suggests that BGE-M3 is better suited to our task.

\begin{figure}
    \centering
    \includegraphics[width=0.8\linewidth]{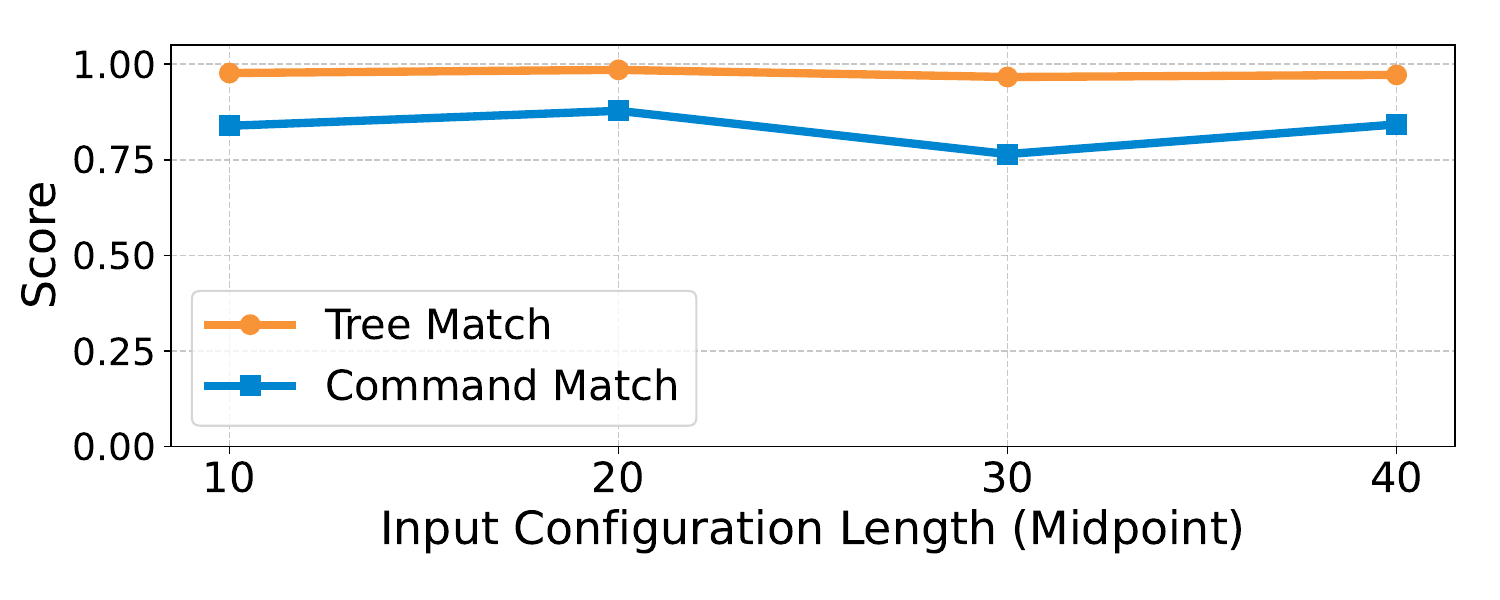}
    \caption{\revised{\method{} performance across varying configuration lengths (\method{} with DeepSeek-V3). Each x-axis value represents the midpoint of a length interval, where the source device configurations fall within the range $[x{-}5,\,x{+}5)$.}}
    \label{fig:config_length_performance}
\end{figure}

\section{\revised{Performance across configuration lengths}}\label{sec:length}

\revised{To analyze \method{}'s performance variation with input length, we group source configurations by length and evaluate performance across these groups.
Due to data availability, our dataset includes source device configuration files ranging from 10 to 40 lines, each representing a semantically complete unit.
As shown in Fig.~\ref{fig:config_length_performance}, \method{} maintains stable performance across this range, with both Tree Match and Command Match metrics showing no degradation as length increases.
This range also reflects practical translation units, as network engineers often translate modular blocks in a single pass.
For example, core OSPF and BGP configurations typically span only a few dozen lines, while ACL policies may reach several hundred.
Although our experiments are constrained by data availability, \method{}'s architecture is designed with the capacity to scale to much longer inputs.
Its fragment-based architecture enables incremental translation and scales naturally, with the maximum input length primarily bounded by the LLM's context window.
This positions \method{} to handle increasingly complex configurations as model capacity continues to grow.}

\end{document}